\documentclass[twocolumn,showpacs,superscriptaddress,longbibliography,nofootinbib,pra]{revtex4-1}
\usepackage[utf8]{inputenc}
\usepackage{calrsfs}
\usepackage{graphicx}
\usepackage{textcomp}
\usepackage{soul}
\usepackage{amsmath,amssymb}
\usepackage{color}
\usepackage{braket}
\usepackage{xcolor}
\usepackage{float}
\usepackage{physics}
\usepackage{soul}
\usepackage{graphicx}
\usepackage{caption}
\usepackage{amsmath}
\usepackage{amssymb}
\usepackage{bbold}
\usepackage{makecell}
\usepackage{tikz,pgfplots}

\usetikzlibrary{decorations.pathmorphing, arrows.meta, positioning, calc, shadings, shapes.geometric, shadows.blur, fadings, fit, shapes, backgrounds}

\definecolor{gainRed}{RGB}{215, 48, 39}
\definecolor{lossBlue}{RGB}{69, 117, 180}
\definecolor{deepSpace}{RGB}{40, 40, 50}
\definecolor{gridLine}{RGB}{200, 200, 220}
\definecolor{SYKLblue}  {RGB}{ 25,  90, 190}
\definecolor{SYKRteal}  {RGB}{  0, 128,  90}
\definecolor{GWred}     {RGB}{190,  35,  20}
\definecolor{TFDpurp}   {RGB}{120,   0, 170}
\definecolor{COUPamb}   {RGB}{185, 110,   0}
\definecolor{BULKgray}  {RGB}{220, 223, 235}
\definecolor{MSGblue}   {RGB}{ 25,  75, 170}
\definecolor{OUTgreen}  {RGB}{  0, 115,  75}
\definecolor{STEPgray}  {RGB}{ 80,  80,  80}

\tikzset{
  gwave/.style={
    line width=1.3pt, decorate,
    decoration={snake, amplitude=3pt, segment length=8pt, post length=4pt},
    -{Stealth[length=6pt, width=4pt]}
  },
  dblarrow/.style={
    line width=1.4pt, dashed,
    {Stealth[length=5pt,width=3.5pt]}-{Stealth[length=5pt,width=3.5pt]}
  },
  singlearrow/.style={
    line width=1.4pt, dashed,
    -{Stealth[length=5pt,width=3.5pt]}
  },
  proarrow/.style={
    line width=2.0pt,
    -{Stealth[length=7pt, width=5pt]}
  },
  stepbox/.style={
    draw=#1, fill=#1!8, rounded corners=4pt,
    inner sep=5pt, line width=1pt
  }
}

\usepackage{mathtools}
\usepackage{multirow}
\usepackage{subcaption}
\usepackage{url}
\usepackage[colorlinks=true, urlcolor=blue, citecolor=blue, linkcolor=blue]{hyperref}
\begin{document}
\title{Gravitational Wave-Inspired Scrambling Delay in SYK Wormhole Teleportation}
\author{Sudhanva Joshi }
\email[]{sudhanvajoshi.rs.phy24@itbhu.ac.in}
\affiliation{Department of Physics, Indian Institute of Technology (Banaras Hindu University), Varanasi - 221005, India}

\author{Sunil Kumar Mishra}
\email[]{sunilkm.app@iitbhu.ac.in}
\affiliation{Department of Physics, Indian Institute of Technology (Banaras Hindu University), Varanasi - 221005, India}

\begin{abstract}
Traversable wormhole teleportation in the Sachdev-Ye-Kitaev (SYK) model links quantum channel integrity directly to black hole interior dynamics: the teleportation fidelity is a boundary-accessible probe of the many-body scrambling that underlies the holographic wormhole. We subject the SYK boundary to a gravitational-wave (GW)-inspired periodic Floquet deformation, constructed from the lowest-dimension bilinear sector of the SYK$_4$ interaction and motivated by the JT-gravity holographic dictionary, and characterize the channel response via exact numerical time evolution with disorder averaging at $\beta J = 2$. The drive produces a coherent, frequency-selective fidelity suppression, validated by re-optimization under drive to separate genuine effects from calibration mismatch with four main results: (i)~two distinct amplitude regimes separated near $\varepsilon \sim J$, a perturbative sensing regime and a non-perturbative strong-drive regime; (ii)~the channel acts as a natural low-pass filter, most sensitive at $\omega \lesssim \beta^{-1}$ with monotone recovery above the thermal scale; (iii)~an inspiral chirp drive displaces the fidelity peak to later readout times (by $\Delta t_{\rm scr}^{(\rm fid)} = +0.11\,J^{-1}$, a sub-grid displacement), and an out-of-time-order correlator (OTOC) diagnostic under a monochromatic drive independently resolves a scrambling delay of the same sign ($\Delta t_{\rm scr}^{(\rm OTOC)} = +0.20\,J^{-1}$ at $\varepsilon=0.20\,J$, $+0.85\,J^{-1}$ at $\varepsilon=0.50\,J$), with the delay growing monotonically with drive amplitude; (iv) the drive-induced fidelity suppression is non-zero and shows no systematic decrease across $N \in \{10,\,12,\,14,\,16\}$ Majorana modes, and is resolved above the disorder noise at three of the four system sizes. These results establish that holographic teleportation channels degrade gracefully and can be diagnosed via GW-inspired metric-like boundary deformations, with direct implications for near-term quantum-processor implementations of traversable wormhole protocols.
\end{abstract}

\maketitle
\section{Introduction}
The discovery of gravitational waves (GW) by the Laser Interferometer Gravitational Wave Observatory (LIGO) \cite{abbott2016observation} has opened a new observational window onto the strong-field, dynamical regime of General Relativity. In parallel, the past decade has seen rapid progress in understanding quantum gravity through holographic duality, in which strongly coupled quantum systems on the boundary of an asymptotically anti-de-Sitter spacetime are dual to gravitational theories in the bulk \cite{maldacena1999large,gubser1998gauge,klebanov1999ads,bousso2002holographic,klebanov2001tasi, hubeny2015ads}. \\
A natural question at the intersection of these two developments is whether the characteristic signatures of a gravitational wave, such as its frequency, amplitude, and temporal structure, can be imprinted on, and read out from, purely quantum-information-theoretic observables defined on the holographic boundary \cite{chakrabortty2023holographic}. \\
A natural and particularly compelling setting for this question is the Sachdev-Ye-Kitaev (SYK) model \cite{sachdev1993gapless}, a $0{+}1$ - dimensional system of $N$ Majorana fermions with random all-to-all interactions, which saturates the Maldacena-Shenker-Stanford (MSS) bound on quantum chaos \cite{maldacena2016bound} and admits a low-energy dual description in terms of Jackiw-Teitelboim (JT) gravity \cite{jackiw1985lower,teitelboim1983gravitation,mertens2023solvable,maldacena2016remarks}. The SYK model is both analytically tractable in the large-$N$ limit and numerically accessible at finite-$N$ \cite{lunkin2018sachdev,rosenhaus2019introduction}, making it an ideal testbed for holographic ideas in controllable quantum many-body systems.\\
A key holographic construction is the traversable wormhole \cite{gao2017traversable,gao2021traversable}, which elucidates that the coupling between the two boundaries of an eternal AdS black hole via a double trace deformation creates a traversable channel through the bulk. The SYK realization of this mechanism, analyzed by \cite{maldacena2018eternal}, leads to a concrete quantum teleportation protocol wherein a signal injected into the left boundary traverses the wormhole and emerges at the right boundary with measurable fidelity, provided the two sides share thermofield double (TFD) entanglement at inverse temperature $\beta$. This Wormhole-inspired teleportation protocol \cite{brown2023quantum,nezami2023quantum,schuster2022many, Shapoval2023towardsquantum} has recently been implemented on a quantum processor \cite{jafferis2022traversable}, establishing SYK-based teleportation as a particularly realizable quantum information primitive. \\
Operationally, traversable wormhole teleportation defines a concrete quantum communication task: an unknown qubit state is encoded into simple boundary operators on one side of a two-sided many-body system and is recovered on the other side after a prescribed sequence of time evolutions and a brief inter-boundary coupling. The performance of this channel is quantified by a teleportation fidelity $\mathcal{F}$ measured entirely from boundary data; in our convention, $\mathcal{F}=0.25$ is the classical benchmark for the protocol, so $\mathcal{F}>0.25$ certifies nontrivial quantum-information transfer within this decoding task.\\
Despite this progress, the sensitivity of the WITP to dynamical perturbations of the bulk geometry has received little to no systematic attention. There arises an interesting question: How robust is a holographic quantum channel to GW-like perturbations, and what does the answer reveal about the scrambling dynamics that underpin it? In the simplest $0{+}1$-d SYK/JT duality, propagating gravitational waves do not exist as independent degrees of freedom \cite{engelsoy2016investigation}; however, one can model the boundary imprint of a passing gravitational wave as a time-dependent bilinear Floquet deformation of the SYK Hamiltonian, coupling the GW strain amplitude to the boundary stress tensor. In higher-dimensional holographic theories, metric perturbations source the boundary stress tensor; here we use the lowest-dimension fermion-bilinear sector as an SYK-appropriate analog of a metric-like strain channel. This construction is the natural boundary image of a metric perturbation sourced by the stress-energy of the dual geometry \cite{gao2017traversable,maldacena2018eternal}, and defines a well-posed and computable Floquet perturbation of the SYK-WITP channel whose physical consequences have not previously been studied.  \\
We emphasize at the outset that the deformation studied here is \emph{GW-inspired} rather than a controlled holographic model of a propagating gravitational wave. In the $0{+}1$-dimensional SYK/JT duality, there are no propagating graviton modes, and the connection to astrophysical GW signals is one of waveform structure and physical motivation rather than of dynamical equivalence. The question we address is self-contained and well posed: how does a time-dependent bilinear deformation of the SYK Hamiltonian affect the wormhole teleportation channel, and what do standard quantum-information observables reveal about the underlying scrambling dynamics? All quantitative claims in this paper concern that question. \\
In this paper, we carry out a systematic numerical investigation of this GW-inspired boundary strain on the SYK-WITP channel at finite temperature ($\beta J=2$) and finite system size ($N=10$--$16$) Majorana modes. We model the GW drive as a time-periodic Floquet bilinear deformation $\varepsilon\,h(t)\, H_{\rm strain}$ of the SYK Hamiltonian, where $\varepsilon$ is the dimensionless strain amplitude and $h(t)$ encodes the waveform. Using exact diagonalization, Lie-Trotter time evolution, and disorder averaging, we compute the teleportation fidelity, its frequency-resolved spectroscopy, and independent out-of-time-order correlator (OTOC) diagnostics of the underlying scrambling dynamics.\\ 
Our main results are as follows. (i) The teleportation fidelity is robust at weak-to-moderate strain (small $\varepsilon$) but undergoes systematic suppression as the drive amplitude increases, with two distinct operating regimes separated near $\varepsilon \sim J$: a sensing regime in which protocol degradation is a genuine channel effect, and a strong drive protocol shift regime in which the GW drive shifts the optimal operating point of the channel itself. (ii) Frequency scans reveal maximum sensitivity in the quasi-static regime $\omega \lesssim \omega_T = \beta^{-1}$, with monotone recovery above the thermal crossover; the sensitivity is low-pass rather than narrow-band, with the suppression falling into the disorder noise floor near $\omega \approx 1\,J$, well below the chaos scale $\omega_L = 2\pi\beta^{-1}$, which therefore plays no empirical role in the measured response. (iii) An inspiral chirp drive displaces the teleportation peak by $\Delta t_{\rm scr}^{(\rm fid)} = +0.11\, J^{-1}$; this displacement is sub-grid and indicates the direction of the effect rather than resolving its magnitude. An OTOC scrambling-time shift $\Delta t_{\rm scr}^{(\rm OTOC)} = +0.20\, J^{-1}$ at fixed $\varepsilon=0.20 \, J$ independently resolves a delay of the same sign, ruling out a pure protocol-calibration artifact. (iv) The drive-induced fidelity suppression $\Delta\mathcal{F}^*$ is non-zero and shows no systematic decrease over the accessible finite-size range $N\in\{10,12,14,16\}$, and is resolved well above the disorder noise at three of the four system sizes, motivating its interpretation as a robust many-body effect rather than a small-Hilbert-space artifact. 
The baseline Wormhole-Inspired Teleportation Protocol, including the Thermofield Double construction, numerical optimization of the coupling parameters $(g^{*}, t^{*})$, and finite-size fidelity characterization, was established in our recent work \cite{joshi2025sachdev}. Here, we elevate that baseline into a dynamical channel-diagnostics problem: we ask how a time-dependent deformation modifies both channel fidelity and scrambling observables, and we validate the mechanism with independent tests (re-optimization and OTOCs). The present manuscript introduces four physically new contributions that are entirely absent from that work: (i) the construction and normalization of a GW-inspired bilinear Floquet strain operator $H^\alpha_{\rm strain}$ as a boundary analog of a metric perturbation; (ii) a systematic amplitude and frequency spectroscopy of the WITP fidelity under this drive; (iii) the identification and dual OTOC verification of a GW-induced many-body scrambling delay; and (iv) the re-optimization analysis that isolates genuine holographic channel degradation from protocol calibration mismatch.

Taken together, these results establish wormhole-inspired teleportation as a calibrated probe of time-dependent (GW-inspired) boundary deformations in a strongly scrambling quantum channel, with implications for near-term implementations of traversable-wormhole circuits and for diagnosing drive-induced changes in scrambling using standard quantum-information observables.

\section{Setup and Model} \label{sec:model}
The system consists of two copies of the SYK model, labeled $L$ and $R$, prepared in a thermofield double state \cite{su2021variational} and connected by a coupling $g$ which makes the wormhole traversable at a single moment. A gravitational-wave-inspired Floquet deformation is applied to the boundary dynamics; in the amplitude and frequency scans, it acts on both boundaries, whereas in the chirp experiment, it is applied only to the right boundary during readout (see below). The full geometry is illustrated in Fig.~(\ref{fig:setup1}).

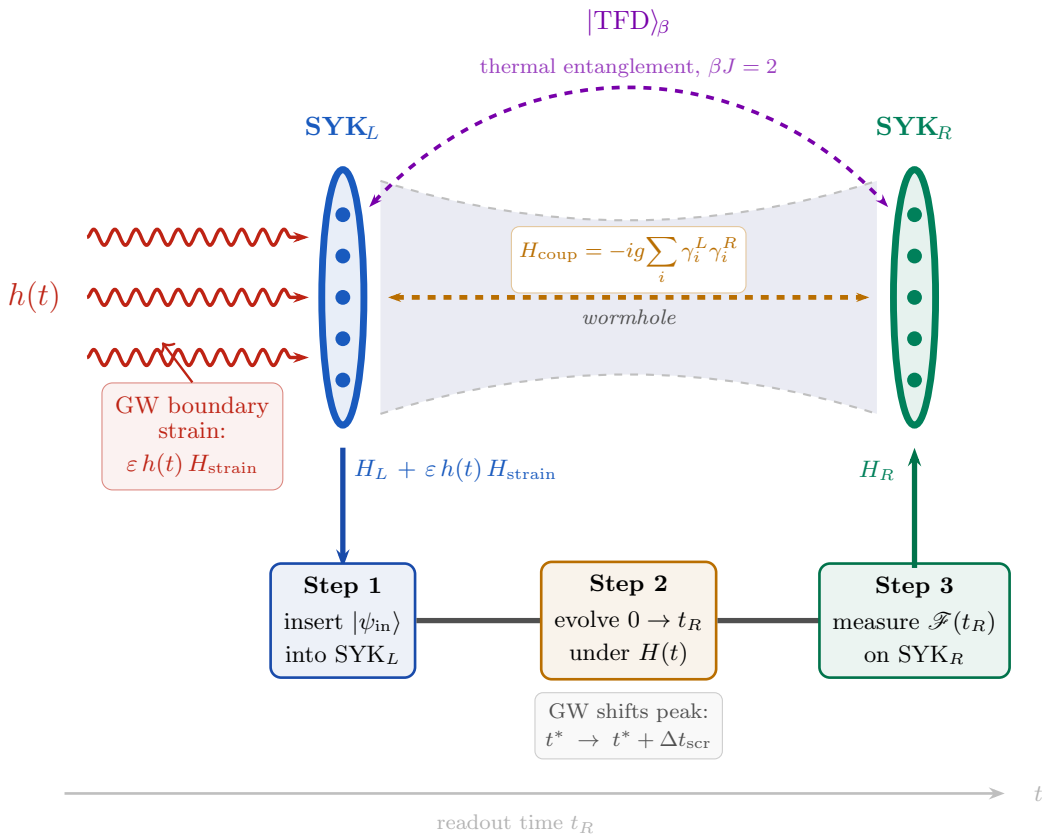
\begin{figure*}
\centering
\begin{tikzpicture}[scale=1.0, every node/.style={font=\small}]

\begin{scope}[on background layer]
  \fill[BULKgray, opacity=0.60]
    (-3.3,  1.55) .. controls (-1.1,  0.85) and ( 1.1,  0.85) .. ( 3.3,  1.55)
    -- ( 3.3, -1.55) .. controls ( 1.1, -0.85) and (-1.1, -0.85) .. (-3.3, -1.55)
    -- cycle;
\end{scope}

\draw[SYKLblue, line width=2.6pt, fill=SYKLblue!10]
  (-3.8, 0) ellipse (0.28 and 1.70);

\foreach \y in {-1.1, -0.55, 0.0, 0.55, 1.1}
  \fill[SYKLblue] (-3.8, \y) circle (2.8pt);

\node[SYKLblue, font=\normalsize\bfseries] at (-3.8, 2.25) {SYK$_{\!L}$};

\node[SYKLblue, align=center, text width=4.2cm, font=\small]
  at (-2.3, -2.30)
  {$H_L + \varepsilon\,h(t)\,H_{\mathrm{strain}}$};

\draw[SYKRteal, line width=2.6pt, fill=SYKRteal!10]
  (3.8, 0) ellipse (0.28 and 1.70);

\foreach \y in {-1.1, -0.55, 0.0, 0.55, 1.1}
  \fill[SYKRteal] (3.8, \y) circle (2.8pt);

\node[SYKRteal, font=\normalsize\bfseries] at (3.8, 2.25) {SYK$_{\!R}$};

\node[SYKRteal, align=center, font=\small]
  at (3.3, -2.30) {$H_R$};

\draw[gray!50, dashed, line width=0.8pt]
  (-3.3, 1.55) .. controls (-1.1, 0.85) and (1.1, 0.85) .. (3.3, 1.55);
\draw[gray!50, dashed, line width=0.8pt]
  (-3.3,-1.55) .. controls (-1.1,-0.85) and (1.1,-0.85) .. (3.3,-1.55);

\node[gray!60!black, font=\footnotesize\itshape] at (0, 0.18) {bulk};
\node[gray!60!black, font=\footnotesize\itshape] at (0,-0.25) {wormhole};

\draw[TFDpurp, dblarrow]
  (-3.45, 1.20) .. controls (-1.5, 3.30) and (1.5, 3.30) .. (3.45, 1.20);

\node[TFDpurp, font=\normalsize, fill=white, inner sep=2pt]
  at (0, 3.65) {$|\mathrm{TFD}\rangle_{\!\beta}$};

\node[TFDpurp!75, font=\footnotesize]
  at (0, 3.05) {thermal entanglement, $\beta J = 2$};

\draw[COUPamb, line width=1.8pt,
      {Stealth[length=5pt,width=3.5pt]}-{Stealth[length=5pt,width=3.5pt]},
      dashed]
  (-3.2, 0) -- (3.2, 0);

\node[COUPamb, fill=white, draw=COUPamb!40, rounded corners=3pt,
      inner sep=3pt, font=\footnotesize]
  at (0, 0.50)
  {$H_{\mathrm{coup}} = -ig\!\displaystyle\sum_i \gamma_i^L \gamma_i^R$};

\foreach \y in {0.80, 0.0, -0.80}
  \draw[GWred, gwave] (-7.20, \y) -- (-4.25, \y);

\node[GWred, font=\large\bfseries] at (-7.90, 0) {$h(t)$};

\node[GWred, draw=GWred!55, fill=GWred!6, rounded corners=4pt,
      inner sep=5pt, font=\small, align=center]
  (gwbox) at (-5.80, -1.85)
  {GW boundary\\[-1pt]strain:\\[2pt]
   $\varepsilon\,h(t)\,H_{\mathrm{strain}}$};

\draw[GWred, ->, line width=0.9pt]
  (gwbox.north) -- (-6.20, -0.60);


\draw[MSGblue, proarrow]
  (-3.8, -2.00) -- (-3.8, -3.60);

\node[stepbox=MSGblue, font=\small, align=center]
  (S1) at (-3.8, -4.30)
  {\textbf{Step 1}\\[3pt]
   insert $|\psi_{\mathrm{in}}\rangle$\\[2pt]
   into SYK$_L$};

\draw[STEPgray, proarrow]
  (S1.east) -- ++(1.40, 0) |- (0, -4.30);

\node[stepbox=COUPamb, font=\small, align=center]
  (S2) at (0, -4.30)
  {\textbf{Step 2}\\[3pt]
   evolve $0 \to t_R$\\[2pt]
   under $H(t)$};

\draw[STEPgray, proarrow]
  (S2.east) -- ++(1.40, 0) |- (3.8, -4.30);

\node[stepbox=OUTgreen, font=\small, align=center]
  (S3) at (3.8, -4.30)
  {\textbf{Step 3}\\[3pt]
   measure $\mathcal{F}(t_R)$\\[2pt]
   on SYK$_R$};

\draw[OUTgreen, proarrow]
  (3.8, -3.60) -- (3.8, -2.00);

\node[STEPgray, font=\footnotesize, align=center,
      draw=gray!40, fill=gray!5, rounded corners=3pt, inner sep=4pt]
  at (0, -5.70)
  {GW shifts peak:\\[1pt]
   $t^* \;\to\; t^* + \Delta t_{\mathrm{scr}}$};

\draw[gray!50, -{Stealth[length=6pt]}, line width=1.0pt]
  (-7.5, -6.60) -- (5.0, -6.60);
\node[gray!60, font=\small] at (5.45, -6.60) {$t$};
\node[gray!60, font=\footnotesize] at (-1.5, -7.00)
  {readout time $t_R$};

\end{tikzpicture}
\caption{Schematic of the GW-perturbed wormhole teleportation protocol(GW-WITP)}
\label{fig:setup1}
\end{figure*}

Each boundary is described by an $\rm SYK_4$ Hamiltonian 
\begin{equation}
  H_{\alpha} = -\frac{1}{4!}\sum_{i<j<k<l}^{N}
    J_{ijkl}\,\gamma_i^{\alpha}\gamma_j^{\alpha}
    \gamma_k^{\alpha}\gamma_l^{\alpha},
  \qquad \alpha \in \{L, R\},
  \label{eq:HSYK}
\end{equation}
where $N$ Majorana fermion operators $\{\gamma_i^\alpha\}$ satisfy $\{\gamma_i,\gamma_j\} = 2\delta_{ij}$, and the all-to-all random couplings are drawn independently from 
\begin{equation}
  J_{ijkl} \;\sim\; \mathcal{N}\!\left(0,\;\frac{6J^2}{N^3}\right).
  \label{eq:Jdist}
\end{equation}
The variance $6J^2/N^3$ is the standard large-$N$ normalization ensuring extensive ground-state energy \cite{kitaev2015simple}, and we express all energies in units of $J$. Both boundaries share the same disorder realization $\{J_{ijkl}\}$; the right boundary Hamiltonian is obtained by the replacement $\gamma_i^R \to -i\gamma_i^R$, implementing the charge-conjugation symmetry required by the thermofield double construction \cite{haenel2021traversable}. \\
The Majorana operators are represented by the Jordan-Wigner (JW) transformation on a chain of $N/2$ complex fermions, 
\begin{align}
  f_k &= \Bigl(\prod_{m<k}\sigma_z^{(m)}\Bigr)\sigma_-^{(k)}, \\
  \gamma_{2k-1} &= f_k + f_k^\dagger, \qquad
  \gamma_{2k} = -i(f_k - f_k^\dagger),
  \label{eq:splitfermion}
\end{align}
yielding a single-boundary Hilbert space of dimension $d = 2^{N/2}$ ($d = 64$ for $N=12$; $d=128$ for $N=14$). Unless stated otherwise, physical observables are averaged over $N_{\rm avg} = 20$ independent disorder realizations. Two analyses use different ensembles because of their computational cost: the re-optimization study of Sec.~\ref{sec:reopt} and the OTOC diagnostics of Sec.~\ref{sec:chirp} use $N_{\rm avg}=5$, while the system-size scan of Sec.~\ref{sec:scaling} uses $N_{\rm avg}=50$. Error bars and shaded regions throughout denote the disorder standard error
$\sigma_{\rm dis}/\sqrt{N_{\rm avg}}$ for the ensemble in use. \\

The two-sided system is initialized in the thermofield double (TFD) state
\begin{equation}
|\mathrm{TFD}\rangle_\beta=\frac{1}{\sqrt{Z}}\sum_n e^{-\beta E_n/2}\,
|n\rangle_L\otimes |n\rangle_R.
\end{equation}
Equivalently, following ~\cite{qi2020coupled,lin2022bulk}, we write
\begin{equation}
|\mathrm{TFD}\rangle_\beta=
\frac{e^{-\beta H_L/2}\,|I\rangle}{\|e^{-\beta H_L/2}\,|I\rangle\|},
\end{equation}
where $|I\rangle$ is the maximally entangled ``infinite-temperature TFD'' satisfying $c_i|I\rangle=0$ for the nonlocal fermions $c_i=(\gamma_i^L+i\gamma_i^R)/2$. This choice ensures $(H_L-H_R)|I\rangle=0$ and produces the standard TFD purifying the Gibbs state $\rho_L=e^{-\beta H_L}/Z$ and its entanglement structure is the boundary dual of the two-sided eternal AdS black hole \cite{maldacena2003eternal}. \\
The full initial state of the protocol combines the TFD state with a Bell pair spanning the message qubit $m$ and a reference ancilla $a$, 
\begin{equation}
  |\Psi_0\rangle
  = |\mathrm{\Phi^+}\rangle_{m,a}
    \otimes|\mathrm{TFD}\rangle_{L,R},\quad
  |\Phi^{+}\rangle
  = \frac{|00\rangle+|11\rangle}{\sqrt{2}},
  \label{eq:Psi0}
\end{equation}
so that the reference ancilla retains a copy of the message for fidelity evaluation at the end of the protocol. All results use $\beta J=2$, placing the system in a low-temperature holographic regime where SYK fast-scrambling is active and the large-$N$ Lyapunov exponent $\lambda_L = 2\pi/\beta$ saturates the chaos bound in the large-$N$ conformal limit \cite{maldacena2016bound}. \\

The full teleportation protocol then proceeds in three steps; starting from $\ket{\Psi_0}$, the left boundary is evolved backward in time from $t=0$ to $t=-t^*$ under the GW-perturbed Hamiltonian $H_L(-t)$. At $t=-t^*$, the message qubit is encoded into the left boundary via the Pauli-channel map 
\begin{equation}
  \rho \;\mapsto\; \frac{1}{2}\sum_{\mu=0}^{3}
    (\sigma_\mu^L \otimes \sigma_\mu^m)\,
    \rho\,
    (\sigma_\mu^L \otimes \sigma_\mu^m)^\dagger,
  \label{eq:insert}
\end{equation}
where $\sigma_0 = \mathbf{1}$, $\sigma_{1,2,3} = \sigma_{x,y,z}$
act on the first qubit of the left boundary and on the message
qubit respectively. This Eq.~(\ref{eq:insert}) is the Hayden-Preskill encoding \cite{hayden2007black}, the quantum circuit analog of throwing the particle through the left AdS boundary \cite{gao2017traversable}. \\
The next step is to evolve the left boundary from $ t =- t^*$ to $ t = 0$ under $H_L(t)$. At $t=0$, the traversable wormhole interaction or coupling is applied as the instantaneous unitary 
\begin{equation}
  U_g = \exp\!\left(ig\sum_{i=2}^{N-1} n_i\right),
  \quad
  n_i = c_i^\dagger c_i,\quad
  c_i = \frac{\gamma_i^L + i\gamma_i^R}{2},
  \label{eq:Ug}
\end{equation}
where $c_i$ is the non-local fermion straddling both boundaries, and the sum runs from $i = 2$ to $N-1$ because the first two Majorana modes ($i = 0, 1$) are reserved for the left-boundary qubit used in the message encoding. In the code, Majorana arrays follow standard 0-based indexing $\gamma_0, \ldots, \gamma_{N-1}$. $U_g$ implements the double-trace deformation \cite{gao2017traversable} at the level of SYK boundary theory \cite{qi2020coupled}, rendering the bulk wormhole traversable and opening a transmission window centered at $t_R \approx t^*$ on the right boundary. \\
Following $U_g$, the right boundary evolves from $t=0$ to readout time $t_R$ under $H_R(t)$. The teleportation fidelity is measured as the overlap between the reconstructed right boundary qubit and the reference ancilla,
\begin{equation}
  \mathcal{F}(t_R)
  = \frac{1}{4}\Bigl(
      1
      + \langle\sigma_x^R\sigma_x^a\rangle
      - \langle\sigma_y^R\sigma_y^a\rangle
      + \langle\sigma_z^R\sigma_z^a\rangle
    \Bigr),
  \label{eq:fidelity}
\end{equation}
where expectation values are taken in the state evolved up to $t_R$. A random (classical) guess yields $\mathcal{F} = 1/4$. The optimal parameters $(g^*,t^*)$ are found by maximising $\mathcal{F}$ at $\varepsilon = 0$ over the grid $(g, t_R) \in [8, 28]\,J \times [3, 14]\,J^{-1}$ averaged over three disorder seeds, yielding $g^* = 12.0\,J$, $t^* = 7.0\,J^{-1}$, $\mathcal{F}_0 = 0.626$ for $N=12$, $\beta J = 2$. These values are held fixed across all GW-drive calculations. This ``fixed-calibration'' choice is deliberate: it treats the unperturbed optimum as a reference operating point and allows drive-induced changes in $\mathcal{F}(t_R)$ to be interpreted as channel degradation; we later perform an explicit re-optimization test to separate genuine channel effects from calibration mismatch. \\
In $0{+}1$ dimensions, the SYK/JT duality does not contain propagating gravitational-wave modes as independent bulk fields \cite{jensen2016chaos,maldacena2016bound}. We therefore adopt a gravitational-wave-inspired deformation of the boundary dynamics: a time-dependent bilinear (two-body) Floquet perturbation. In higher-dimensional holographic theories, metric perturbations source the boundary stress tensor; here we use the lowest-dimension fermion bilinears $i\gamma_i^\alpha\gamma_j^\alpha$ as the natural SYK analog of a metric-like strain channel (the derivation of this identification within the JT-gravity holographic dictionary, together with its domain of validity, is given in Appendix~\ref{app:schwarzian}). The full-time dependent Hamiltonian on each boundary is
\begin{equation}
  H_\alpha(t)
  = H_\alpha + \varepsilon\,h(t)\,H_{\rm strain}^\alpha,
  \quad \alpha \in \{L, R\},
  \label{eq:Htotal}
\end{equation}
where the boundary strain operator is
\begin{equation}
  H_{\rm strain}^\alpha
  = \sum_{i<j}^{N}
    \widetilde{J}_{ij}\,
    \bigl(i\gamma_i^\alpha\gamma_j^\alpha\bigr),
  \label{eq:Hstrain}
\end{equation}
and the effective two-body couplings are obtained by partially contracting the four-body SYK tensor over the spectator modes,
\begin{equation}
  \widetilde{J}_{ij}
  = \frac{1}{\dbinom{N-2}{2}}
    \!\!\sum_{\substack{k<l\\[1pt]k,l\,\notin\,\{i,j\}}}
    \!\! J_{ijkl}.
  \label{eq:Jtilde}
\end{equation}
Eqns.~(\ref{eq:Hstrain},\ref{eq:Jtilde}) project the four-body SYK operator algebra onto its bilinear sector. This projection is the finite-$N$ realization of the Hubbard-Stratonovich mean-field decoupling that identifies $H_{\rm strain}^\alpha$ as the dominant bilinear channel obtained from a large-$N$ mean-field decomposition of $H_{\rm SYK}$ (Appendix~\ref{app:HS}). In higher-dimensional holography, a bulk metric perturbation sources the boundary stress tensor; in the $0{+}1$-dimensional SYK/JT setting, the bilinear $i\gamma_i\gamma_j$ is the dominant low dimension-$\tfrac{1}{2}$  operator in the SYK$_4$ spectrum and is dual to a bulk scalar at the  AdS$_2$ Breitenlöhner-Freedman bound, making it the leading-order coupling to any boundary metric source in the large-$N$ conformal regime \cite{gao2017traversable}; the limitations of this statement at the parameters studied here are discussed in Appendix~\ref{app:BF}. The operator is spectrally normalized to $\|H^\alpha_{\rm strain}\|_{\rm sp} = 5\, J$ (five times the SYK coupling scale), so that $\varepsilon$ measures the GW strain amplitude in units of $J$. This choice is a normalization convention that fixes the scale of $\varepsilon$ uniformly across disorder realizations and system sizes; changing the constant factor simply rescales the numerical value of $\varepsilon$ without affecting the qualitative conclusions. We note the convention explicitly: $\varepsilon$ is dimensionless, and the physical energy scale of the perturbation is $\varepsilon\,\|H_{\rm strain}^\alpha\|_{\rm sp} = 5\varepsilon J$. Values are quoted below in the form $\varepsilon = 0.20\, J$ as a reminder of this scale, and should be read as $\varepsilon = 0.20$ in units where $J=1$; the crossover written as $\varepsilon \sim J$ is likewise the statement $\varepsilon \sim 1$.\\ 
The deformation is applied differently depending on the experiment. For amplitude scan and frequency spectroscopy (Secs.~\ref{sec:amplitude}--\ref{sec:spectroscopy}), both boundaries carry the strain: the left boundary under $H_L + \varepsilon h(t) H_{\rm strain}^L$ during Steps~1 and~2, and the right boundary under $H_R + \varepsilon h(t) H_{\rm strain}^R$ during Step~3. For the inspiral chirp experiment (Sec.~\ref{sec:chirp}), the left-side preparation is performed under the unperturbed $H_L$ and the GW strain $\varepsilon h(t) H_{\rm strain}^R$ is applied exclusively on the right boundary during Step~3. 
This asymmetric protocol isolates the GW's effect on the readout dynamics independently of perturbations to the message-insertion step, providing a cleaner signature of scrambling delay in the right-side fidelity profile. Because the bilateral protocol applies the strain to both boundaries simultaneously while the chirp experiment applies it to the right boundary only, the bilateral spectroscopy provides an upper bound on the single-boundary response; a unilateral frequency scan at fixed $\varepsilon$ would yield a proportionally smaller $\Delta\mathcal{F}(\omega)$.\\
Two waveform models are used for the envelope $h(t)$ in Eq.~(\ref{eq:Htotal}). Monochromatic drive 
\begin{equation}
  h(t) = \cos(\omega t),
  \label{eq:mono}
\end{equation}
is employed in the amplitude scan and frequency spectroscopy with $\omega \in [0, 4\,J]$. Two physically distinct reference scales characterize the frequency response:

\begin{align}
\omega_T &= \beta^{-1} = 0.50\,J
    \quad\text{(quasi-static crossover)},\\
\omega_L &= 2\pi\beta^{-1} = 3.14\,J
    \quad\text{(MSS bound scale).}
\end{align}

These two scales are not independent: at fixed $\beta$ they differ only by the fixed factor $\omega_L/\omega_T = 2\pi \approx 6.3$. The system therefore admits only two genuinely separated asymptotic regimes, $\omega \ll \omega_T$ and $\omega \gg \omega_T$, with no distinct intermediate regime; accordingly we do not treat $[\omega_T,\omega_L]$ as a spectral window. The single controlling scale is the thermal correlation rate $\omega_T = \beta^{-1}$. For $\omega \ll \omega_T$ the drive is quasi-static over the thermal correlation time and acts as a slow renormalization of the effective couplings, giving maximal fidelity suppression; for $\omega \gtrsim \omega_T$ it averages out within the support of the strain correlator and the suppression falls away. The scale $\omega_L = 2\pi/\beta$ coincides with the large-$N$ Lyapunov exponent $\lambda_L = 2\pi/\beta$ saturating the Maldacena-Shenker-Stanford bound \cite{maldacena2016bound}, and is quoted here only as a reference chaos timescale. As shown in Sec.~\ref{sec:spectroscopy}, the measured suppression has already fallen into the disorder noise floor well below $\omega_L$, so $\omega_L$ plays no empirical role in the observed response and does not delimit a detection band.\\
To model the frequency-sweeping signal of a compact binary inspiral \cite{allen2012findchirp}, we use
\begin{equation}
h(t) = A(t)\,\cos\!\left[
    \omega_T t + \tfrac{1}{2}\dot\omega\,t^2
  \right],
  \quad \dot\omega = \tfrac{5}{7}\,J^{2} \simeq 0.714\,J^{2},
  \label{eq:chirp}
\end{equation}
where $A(t) = \sin(\pi t/(t^*+0.1))$ and the chirp rate $\dot\omega$ is chosen so that the sweep traverses the thermal crossover and extends above the chaos scale within a single teleportation run. The envelope $A(t)$ is unit-normalised (peak value 1); the overall strain amplitude $\varepsilon_0 = 0.50\, J$ enters exclusively through the Hamiltonian prefactor in Eq.~(\ref{eq:Htotal}), so that the peak Hamiltonian perturbation is $\varepsilon_0 \lVert H_{\rm strain}^R \rVert_{\rm sp} = 2.5\, J$. $A(t)$ is a smooth on/off envelope, and the instantaneous frequency
\begin{equation}
  \omega_{\rm inst}(t) = \omega_T + \dot{\omega}\,t
  \label{eq:omega_inst_def}
\end{equation}
sweeps linearly from $\omega_T = 0.50\,J$ at $t=0$ to $\omega_T + \dot\omega\,t^* = 5.5\,J$ at $t = t^*$, traversing the full thermal-crossover region and extending above $\omega_L$ during a single teleportation run. \\
Time evolution under the time-dependent Hamiltonian $H_\alpha(t)=H_\alpha+\varepsilon h(t)H_{\rm strain}^\alpha$ is implemented by a piecewise-constant approximation on steps of size $\delta t$,
\begin{equation}
U(t+\delta t,t)\approx \exp\!\left[-i\delta t\,\bigl(H_\alpha+\varepsilon h(t)\,
H_{\rm strain}^\alpha\bigr)\right],
\end{equation}
with the sparse exponential action evaluated using \texttt{expm\_multiply} \cite{al2011computing}. Because $H_\alpha(t)$ is time-dependent, the error committed locally at each step is $O(\delta t^{2})$, with leading coefficient proportional to $\varepsilon\,\|[H_\alpha, H_{\rm strain}^\alpha]\|$; accumulating $t/\delta t$ such steps gives a global error $O(\delta t)$ over the protocol window. The second-order Strang splitting has local error $O(\delta t^{3})$ and global error $O(\delta t^{2})$. We verify convergence under step-size halving and cross-check the two integrators against each other \cite{strang1968construction,mclachlan2002splitting}. \\
The base step is $\delta t = 0.05\,J^{-1}$, reduced adaptively to $\delta t = 0.05/(1+0.5\varepsilon)\,J^{-1}$ at large amplitudes ($\varepsilon > 1\,J$). Convergence against step-size halving is verified at six representative parameter points spanning the full physical range, and cross-checked against a second-order midpoint Strang scheme; in every case, the integration error lies below the disorder noise floor $\sigma_{\rm dis}/\sqrt{N_{\rm avg}}$. Full details are given in Appendix~\ref{app:convergence}.

\section{Amplitude dependence and sensing regimes} \label{sec:amplitude}

We begin the phenomenological study with the most basic observable: how does the teleportation fidelity depend on the GW strain at fixed frequency? Operationally, $\mathcal{F}(\varepsilon)$ quantifies how a time-dependent deformation degrades the effective quantum channel implemented by the protocol at a fixed calibration point, and whether the channel remains functional relative to the classical benchmark $\mathcal{F}=0.25$. Integrator convergence across the full amplitude range is established in Appendix~\ref{app:convergence}.
\begin{figure}
    \centering
    \includegraphics[width=\linewidth]{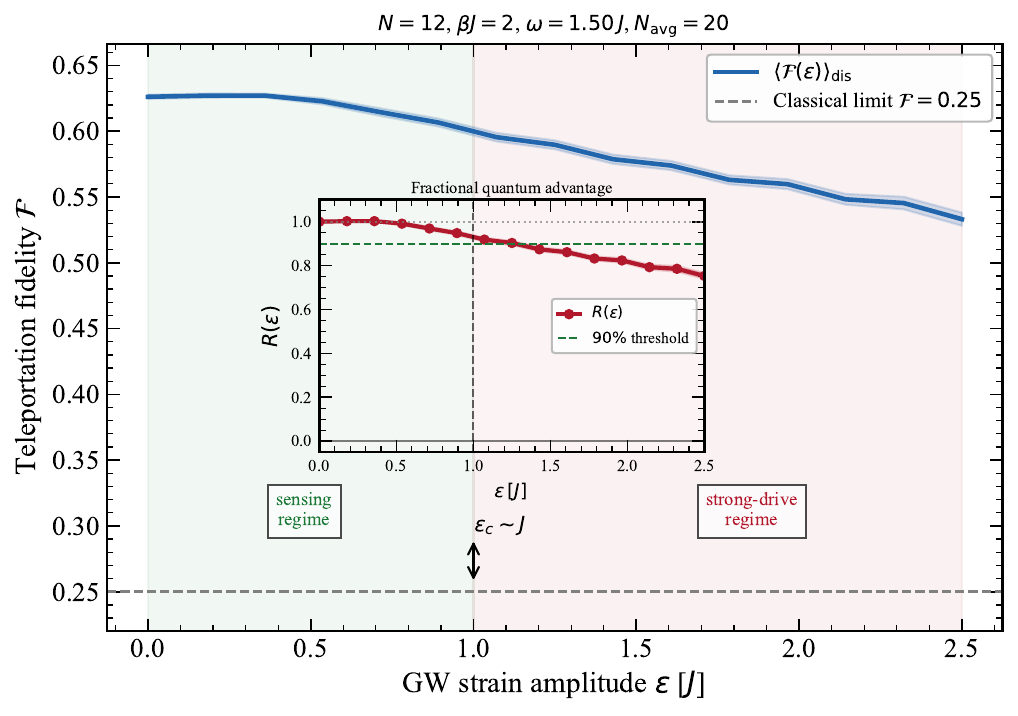}
\caption{\label{fig:amplitude}
Disorder-averaged teleportation fidelity $\mathcal{F}$ versus GW strain amplitude $\varepsilon$ at fixed drive frequency $\omega=1.5\,J$ ($N=12$, $\beta J=2$, $N_{\rm avg}=20$). Shaded bands denote $\pm\sigma_{\rm dis}/\sqrt{N_{\rm avg}}$. The classical benchmark $\mathcal{F}=0.25$ is shown for reference. Across the scanned range $\varepsilon\in[0,2.5J]$, the channel remains well above the classical limit while exhibiting a smooth suppression, monotone within the disorder error bars, with a visibly accelerating rate above the sensing-to-strong-drive crossover near $\varepsilon \sim J$.
\textit{Inset:} Fractional quantum advantage $R(\varepsilon)=(\mathcal{F}(\varepsilon)-0.25)/(\mathcal{F}_0-0.25)$, measuring the fraction of the unperturbed quantum channel capacity retained under the GW drive. $R$ remains above $90\%$ throughout the entire sensing regime $\varepsilon\lesssim J$ and beyond, falling below that threshold only at $\varepsilon \approx 1.3\,J$, well inside the strong-drive regime (dashed vertical line at $\varepsilon = J$). Quantum-advantage teleportation is maintained across the full sensing regime relevant to the results of this paper.}
\end{figure}
Fig.~(\ref{fig:amplitude}) shows the disorder-averaged fidelity $\mathcal{F}(\varepsilon)$ for $\varepsilon \in [0, 2.5\,J]$ at the fixed monochromatic drive $\omega = 1.5\,J$, a representative dynamical (non-quasi-static) drive frequency, well above the thermal scale $\omega_T = 0.50\,J$. Two considerations motivate working at a dynamical rather than a quasi-static frequency. First, for $\omega \ll \omega_T$ the drive is quasi-static over the protocol window and the boundary evolves under an effectively static deformed Hamiltonian $H_\alpha + \varepsilon H_{\rm strain}^\alpha$ (Sec.~\ref{sec:spectroscopy}); that limit therefore probes a renormalized static problem rather than the genuinely time-dependent response of interest here. Second, at $\omega = 1.5\,J$ the amplitude scan extends to $\varepsilon = 2.5\,J$, where the suppression is far above the disorder noise floor, so the two-regime structure reported below is resolved even though the response at $\varepsilon = 0.20\,J$ is not. The adaptive Trotter step $\delta t(\varepsilon) = \max\!\bigl(0.02,\;0.05/(1 + 0.5\varepsilon)\bigr)\, J^{-1}$ is used to maintain integrator error below the disorder noise floor across the full $\varepsilon$ range. \\
For $\varepsilon \lesssim 1\, J$ the fidelity decreases smoothly from the unperturbed baseline $\mathcal{F}_0 = 0.626$, while remaining well above the classical bound of $\mathcal{F}=0.25$ and exhibiting no sudden changes of slope. The disorder-averaged response is consistent with a leading even-power suppression in $\varepsilon$: 
\begin{equation}
  \bigl\langle\mathcal{F}(\varepsilon)\bigr\rangle_{\rm dis}
  \approx \mathcal{F}_0 - \alpha\,\varepsilon^2
    + \mathcal{O}(\varepsilon^4),
  \label{eq:Fpert}
\end{equation}
where $\alpha > 0$ is a non-universal coefficient depending on $\omega$ and $\beta$ (see Appendix.~(\ref{app:scrambling})). The absence of odd powers is a consequence of $\mathbb{Z}_2$ symmetry of Gaussian disorder distribution, $P(\{J_{ijkl}\}) = P(\{-J_{ijkl}\})$: the bilinear couplings $\widetilde{J}_{ij}$ (Eq.~\eqref{eq:Jtilde}) inherit this symmetry \cite{garcia2016spectral}, so $\varepsilon \to -\varepsilon$ corresponds to a statistically equivalent disorder realization, giving $\langle\mathcal{F}(\varepsilon)\rangle_{\rm dis} = \langle\mathcal{F}(-\varepsilon)\rangle_{\rm dis}$.\\
To test the expected even-power scaling, we extract the disorder-averaged OTOC scrambling-time delay $\Delta t_{\rm scr}(\varepsilon)$ at fixed $\omega = 1.5\, J$ for the two drive amplitudes (reported below in Sec.~\ref{sec:chirp} and further analyzed in Appendix~\ref{app:schwarzian}). The measured values,
\begin{equation}
  \Delta t_{\rm scr}(0.20\,J) = +0.20\,J^{-1},
  \qquad
  \Delta t_{\rm scr}(0.50\,J) = +0.85\,J^{-1},
  \label{eq:eps2_test}
\end{equation}
yield the ratio $\Delta t_{\rm scr}(0.50)/\Delta t_{\rm scr}(0.20) \approx 4.25$, corresponding to a two-point power-law exponent $n = \ln(4.25)/\ln(2.5) \approx 1.6$, against the even-power expectation $n = 2$. Two data points cannot determine an exponent with useful precision, and at $\beta J = 2$ and $N=12$ the system lies far from the large-$N$ conformal limit in which the Schwarzian estimate applies. We therefore read this only as evidence for a leading even-power response in $\varepsilon$, and make no claim of quantitative agreement with $\varepsilon^2$; a definitive power-law extraction would require additional amplitudes and larger $N$. The fidelity suppression $\Delta\mathcal{F}(\varepsilon)$ similarly shows an even-in-$\varepsilon$ growth consistent with Eq.~\eqref{eq:Fpert} across the sensing regime (Fig.~\ref{fig:amplitude}). \\
The physical interpretation is that the four-body SYK scrambling is robust against weak bilinear deformations \cite{lunkin2018sachdev}: although the bilinear is a relevant perturbation of the conformal fixed point (Appendix~\ref{app:BF}), the SYK$_4$ infrared behavior is stable against it up to a crossover scale that grows as the perturbation weakens. In the large-$N$ limit, the low-energy sector of the SYK model is governed by a Schwarzian action whose nearly conformal symmetry is only softly broken by bilinear perturbations \cite{maldacena2016bound,jensen2016chaos,mertens2023solvable,das2020near}. A weak strain $\varepsilon H_{\rm strain}^\alpha$ is expected to shift the effective scrambling time by $\Delta t_{\rm scr} \propto \varepsilon^2$, analogous to a marginal deformation of the $\mathrm{AdS}_2$ boundary conditions, without destroying the operational teleportation channel in the weak-drive regime. \cite{kourkoulou2017pure,garcia2021sparse}. The decrease in fidelity is accordingly a small, perturbatively interpretable probe of the GW drive. \\
We confirm the genuine character of this suppression in Sec.~\ref{sec:reopt}: independently re-optimizing the protocol parameters $(g^*, t^*)$ at each $\varepsilon$ recovers at most $\lesssim\!2\%$ of the lost fidelity throughout the sensing regime (Fig.~\ref{fig:reopt}, panel~c), establishing that the suppression is a true GW-induced channel effect rather than a protocol-mismatch artifact. 
Above $\varepsilon \sim J$, the rate of fidelity suppression visibly accelerates, though the transition is smoother and monotone rather than a sharp phase transition. The inset of Fig.~(\ref{fig:amplitude}) quantifies this robustness: the fractional quantum advantage $R(\varepsilon) = (\mathcal{F}(\varepsilon)-0.25)/(\mathcal{F}_0-0.25)$ remains above $90\%$ throughout the sensing regime and falls below that threshold only at $\varepsilon \approx 1.3\,J$, well beyond the onset of the strong-drive regime, confirming that the channel degrades gracefully rather than catastrophically.\\
The crossover is expected when the instantaneous perturbation becomes comparable to a typical spectral scale of the unperturbed dynamics. Since we normalize $\|H_{\rm strain}^\alpha\|_{\rm sp}=5J$, the strong-drive regime corresponds to $\varepsilon\,\|H_{\rm strain}^\alpha\|_{\rm sp}$ becoming $O(1)$ compared to the spectral width of $H_\alpha$ for the finite-$N$ system. \\
In practice, this occurs for $\varepsilon=O(1)$, consistent with the observed change in curvature of $\mathcal{F}(\varepsilon)$ near $\varepsilon\sim J$ in Fig.~(\ref{fig:amplitude}). At this scale, the bilinear strain term can no longer be adiabatically eliminated from the Floquet dynamics: its energy is comparable to the many-body level spacing, enabling resonant multi-photon processes analogous to the onset of Floquet heating in periodically driven quantum systems \cite{d2014long,abanin2016theory,ponte2015many}. In the SYK context, the operator algebra separation between the four-body scrambling sector and the bilinear strain sector, which underlies the robustness of the sensing regime, is overcome once $\varepsilon \gtrsim \varepsilon_c$, and the two sectors begin to mix under the drive \cite{garcia2018exact}. \\
For $\varepsilon \gtrsim 1\, J$, the fidelity continues to decrease but remains well above the classical bound $\mathcal{F} = 0.25$ across the full scanned range up to $\varepsilon = 2.5\, J$. The quantum teleportation channel is degraded but not broken. \\
The key distinction from the sensing regime is operational: in the strong-drive regime, the optimal protocol parameters $(g^*, t^*)$, which are calibrated at $\varepsilon = 0$, are no longer optimal under the drive. Reoptimising $(g^*, t^*)$ at each $(\varepsilon, \omega)$ yields an increasing relative gain in fidelity as $\varepsilon$ increases, reaching $r=\mathcal{F}_{\rm reopt}/\mathcal{F}_{\rm fixed}\lesssim 1.15$ at $(\varepsilon,\omega)=(2.0\,J,0.5\,J)$ in the strong quasi-static corner. This non-trivial re-optimization gain indicates that the effective operating point of the wormhole coupling shifts under a strong GW drive: a Floquet renormalization of the traversable wormhole coupling $g^*$, while the residual, unrecoverable fidelity loss remains a genuine GW-induced effect on the scrambling dynamics. 

Physically, in the holographic dual, the strong-drive regime corresponds to a metric perturbation that is large enough to measurably shift the scrambling time and alter the wormhole traversable window, but not large enough to close the throat or destroy the entanglement structure of the TFD state \cite{maldacena2018eternal,berenguer2024floquet}. The two components of the fidelity suppression, which are protocol-independent (genuine GW channel effect) and protocol-dependent (calibration shift), are disentangled quantitatively through reoptimization analysis in Sec.~\ref{sec:reopt}.

\section{GW Frequency Spectroscopy} \label{sec:spectroscopy}
The amplitude scan of Sec.~\ref{sec:amplitude} characterizes the channel response at a single representative frequency. We now determine the frequencies at which the WITP is most sensitive to a GW perturbation. A useful operational interpretation is that the driven holographic teleportation channel behaves as a natural low-pass filter: slow (quasi-static) deformations degrade the calibrated decoding most strongly, while fast oscillations average out over the thermal correlation time.\\
We define the GW-induced fidelity suppression at frequency $\omega$ as
\begin{equation}
  \Delta\mathcal{F}(\omega)
  = \mathcal{F}_0 - \mathcal{F}(\varepsilon,\omega),
  \label{eq:DeltaF}
\end{equation}
where $\mathcal{F}_0=0.626$ is the unperturbed baseline and $\mathcal{F}(\varepsilon,\omega)$ is measured at fixed amplitude $\varepsilon = 0.20\,J$ with the monochromatic drive $h(t) = \cos(\omega t)$ applied to both boundaries. To second order in $\varepsilon$, the disorder-averaged suppression takes the form
\begin{equation}
  \bigl\langle\Delta\mathcal{F}(\omega)
  \bigr\rangle_{\rm dis}
  \approx \varepsilon^2\,\Sigma(\omega)
  + \mathcal{O}(\varepsilon^4),
  \label{eq:DeltaF_pert}
\end{equation}
where $\Sigma(\omega) \geq 0$ is the fidelity susceptibility, which is a frequency-resolved measure of how efficiently the drive at $\omega$ disrupts the scrambling trajectory. It is schematically controlled by the time-integrated, frequency-resolved autocorrelator of the strain operator in the unperturbed TFD state,
\begin{equation}
  \Sigma(\omega)
  \;\propto\;
  \int_0^{t^*}\!\!\!\int_0^{t}
  \cos\!\bigl[\omega(t - t')\bigr]\,
  C_{\rm strain}(t,t')\;dt'\,dt,
  \label{eq:Sigma}
\end{equation}
with $C_{\rm strain}(t,t') = \langle H_{\rm strain}(t)\,H_{\rm strain}(t')\rangle^{\rm conn}_{\rm TFD}$. Equation~(\ref{eq:Sigma}) is a schematic form valid to leading order in $\varepsilon^2$; the proportionality constant depends on the protocol timing and is not needed for the qualitative frequency dependence. The frequency dependence of $\Sigma(\omega)$ is controlled by two scales: the decay timescale of $C_{\rm strain}$, set by the thermal correlation time $\beta$, and the oscillatory averaging of $\cos[\omega(t-t')]$ over the protocol window $t^*$ \cite{roberts2018operator}.

\begin{figure}
    \centering
    \includegraphics[width=\linewidth]{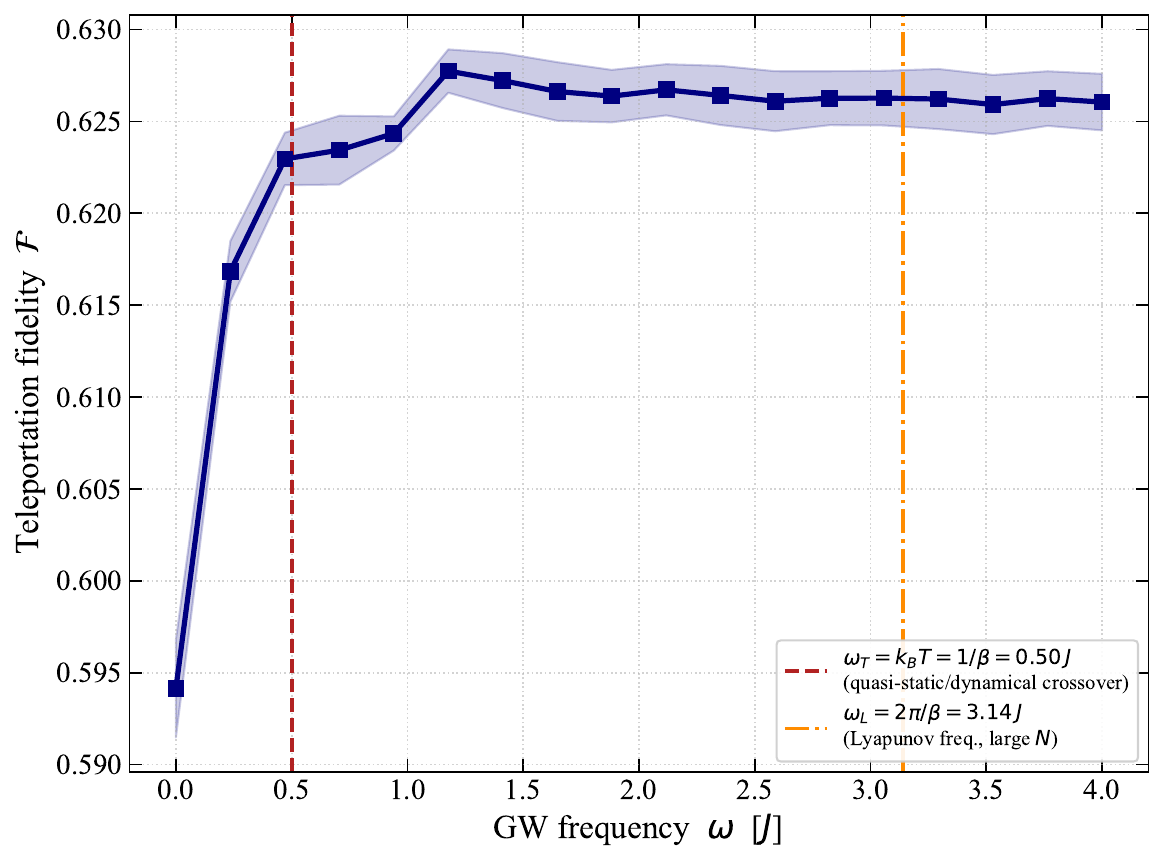}
\caption{\label{fig:spectroscopy}
Teleportation fidelity $\mathcal{F}$ as a function of GW driving frequency $\omega$ at fixed strain amplitude ($N=12$, $\beta J=2$). Two physically distinct frequencies are marked: the thermal scale $\omega_T = \beta^{-1} = 0.50\, J$, which separates the quasi-static and dynamical driving regimes, and the MSS chaos scale $\omega_L = 2\pi/\beta = 3.14\, J$ (large-$N$ chaos timescale). Fidelity suppression is maximum in the quasi-static limit $\omega \to 0$ and decreases with $\omega$, monotonically within the disorder error bars, consistent with a low-pass crossover at $\omega_T$. The sensitivity is not peaked within $[\omega_T, \omega_L]$, and the suppression has already fallen into the disorder noise floor near $\omega \approx 1\,J$, well below $\omega_L$; $\omega_L$ is shown as a reference chaos timescale rather than a band edge. The monotone frequency dependence establishes a low-pass response of the calibrated teleportation channel to GW-inspired boundary deformations.}
\end{figure}

When $\omega\beta \ll 1$, the drive completes far less than one oscillation per thermal correlation time. Expanding $h(t) = \cos(\omega t) = 1 - \tfrac{1}{2}\omega^2 t^2 + \mathcal{O}(\omega^4 t^4)$, the leading constant term dominates: the SYK boundary evolves under the effectively static Hamiltonian $H_\alpha + \varepsilon H_{\rm strain}^\alpha$ throughout the protocol window. \\
The cosine factor in Eq.~(\ref{eq:Sigma}) is approximately unity for all $t,t' \in [0,t^*]$, giving $\Sigma(0) = \Sigma_{\rm max}$, the maximum fidelity susceptibility. The result is the deepest suppression in the entire scan: $\mathcal{F} \to 0.594$ and $\Delta\mathcal{F}_{\rm max} = 0.032$ as $\omega \to 0$, as can be seen from Fig.~(\ref{fig:spectroscopy}). In this regime, the dominant effect is that the drive acts as an effectively static deformation over the protocol window, shifting the effective dynamics away from the $\varepsilon=0$ calibration point and producing the largest suppression.\\
At $\omega\beta \sim1$, the drive begins to oscillate on a time scale comparable to the thermal correlation time. The strain operator $H_{\rm strain}^\alpha \sim i\gamma_i\gamma_j$ is a bosonic (even-parity) bilinear whose two-time correlator $C_{\rm strain}(t,t')$ decays on the scale $\beta$ \cite{eberlein2017quantum}; its spectral weight at finite temperature is appreciable for $\omega \sim \mathcal{O}(\beta^{-1})$ and is governed by the same thermal timescale \cite{engelsoy2016investigation, jensen2016chaos}. The crossover at $\omega_T=\beta^{-1}$  is therefore smooth and monotone: as $\omega$ increases past $\omega_T$, the cosine factor in Eq.~(\ref{eq:Sigma}) begins to oscillate within the support of $C_{\rm strain}$ and the integral decreases continuously.\\
For $\omega \gg \omega_T$ the oscillations in $\cos[\omega(t-t')]$ are rapid relative to the decay of $C_{\rm strain}$, and $\Sigma(\omega)$ is suppressed by oscillatory cancellation:
\begin{equation}
  \Sigma(\omega) \;\xrightarrow{\;\omega\to\infty\;}\; 0,
  \label{eq:SigmaHF}
\end{equation}
so $\Delta\mathcal{F}(\omega) \to 0$ and $\mathcal{F}$ recovers monotonically toward $\mathcal{F}_0$. Within the scanned range $\omega \in [0,4\,J]$ the suppression falls monotonically from $\Delta\mathcal{F} = 0.032$ at $\omega = 0$ to $0.0018$ at $\omega \simeq 0.94\,J$, and is indistinguishable from the disorder noise floor ($\sigma_{\rm dis}/\sqrt{N_{\rm avg}} \simeq 1.6\times10^{-3}$) for all $\omega \gtrsim 1\,J$. The response therefore recovers within roughly two thermal correlation rates of the crossover, and well below the chaos scale $\omega_L = 2\pi/\beta = 3.14\,J$. We retain $\omega_L$ only as a reference timescale (the Lyapunov exponent $\lambda_L \leq 2\pi/\beta$ from the Maldacena-Shenker-Stanford bound~\cite{maldacena2016bound}); it plays no empirical role in the measured response, does not bound a detection band, and does not delimit a spectral window with $\omega_T$. Maximum sensitivity is at the lowest frequencies. \\
The frequency scan therefore characterizes the integrated channel output: the driven WITP behaves as a low-pass filter with a single crossover at the thermal scale $\omega_T$ and no resonant enhancement anywhere in the scanned range. We now turn to a frequency-swept drive, which probes the same physics in the time domain and allows the drive-induced shift of the decoding optimum to be extracted directly.

\section{Chirp Drive and Scrambling-Time Diagnostics} \label{sec:chirp}
\begin{figure}
    \centering
    \includegraphics[width=\linewidth]{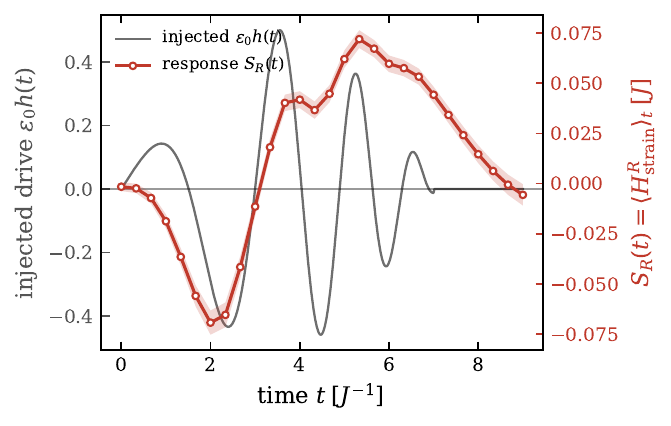}
\caption{\label{fig:detector}
Time-domain comparison of the injected chirp drive $\varepsilon_0 h(t)$ (left axis) and the induced boundary strain response $S_R(t)=\langle H_{\rm strain}^{R}\rangle_t$ [Eq.~(\ref{eq:SR})] (right axis, in units of $J$) during the readout step ($N=12$, $\beta J = 2$, $\varepsilon_0 = 0.50\,J$). The boundary reproduces the temporal structure of the drive with a retarded response that persists for a time of order $\beta$ after the drive terminates at $t = t^*$, and with an envelope that attenuates as the instantaneous frequency sweeps upward. This temporal correlation underlies the fidelity distortion analyzed in this section.}
\end{figure}

\begin{figure}
    \centering
    \includegraphics[width=\linewidth]{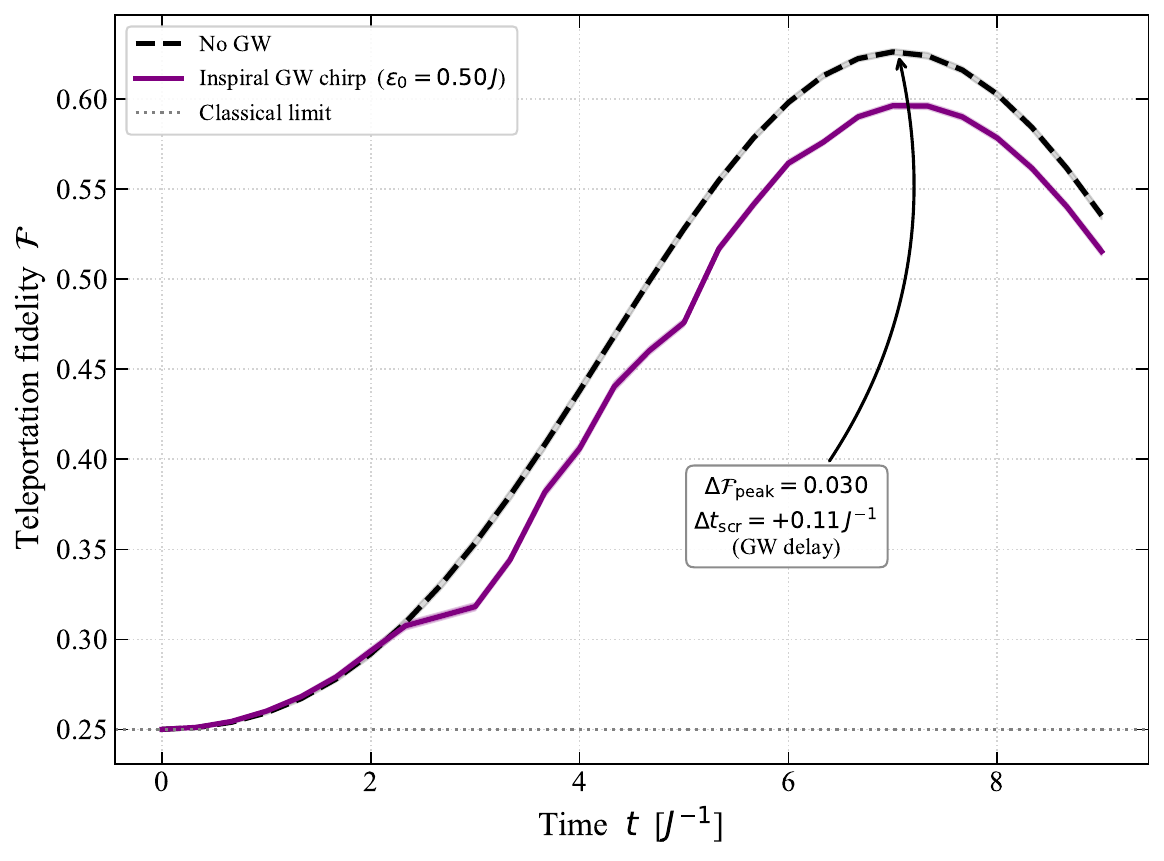}
\caption{\label{fig:chirp}
Teleportation fidelity $\mathcal{F}(t_R)$ vs.\ readout time $t_R$ with (solid, magenta) and without (dashed) an inspiral chirp drive of peak amplitude $\varepsilon_0=0.50\,J$ ($N=12$, $\beta J=2$, $N_{\rm avg}=20$). The drive reduces the peak height by $\Delta\mathcal{F}=0.030$ and displaces the fidelity maximum by $\Delta t_{\rm scr}^{(\rm fid)}=+0.11\,J^{-1}$; the readout grid has spacing $\delta t_R = 0.33\,J^{-1}$, so this displacement is sub-grid and is obtained by parabolic interpolation. The classical teleportation limit (dotted line) is shown for reference. The late-time recovery to the unperturbed baseline establishes the effect as a transient scrambling delay rather than permanent channel disruption. The scrambling-delay interpretation is confirmed by the OTOC diagnostic in Fig.~(\ref{fig:otoc}).}
\end{figure}

\begin{figure*}
    \centering
    \includegraphics[width=\linewidth]{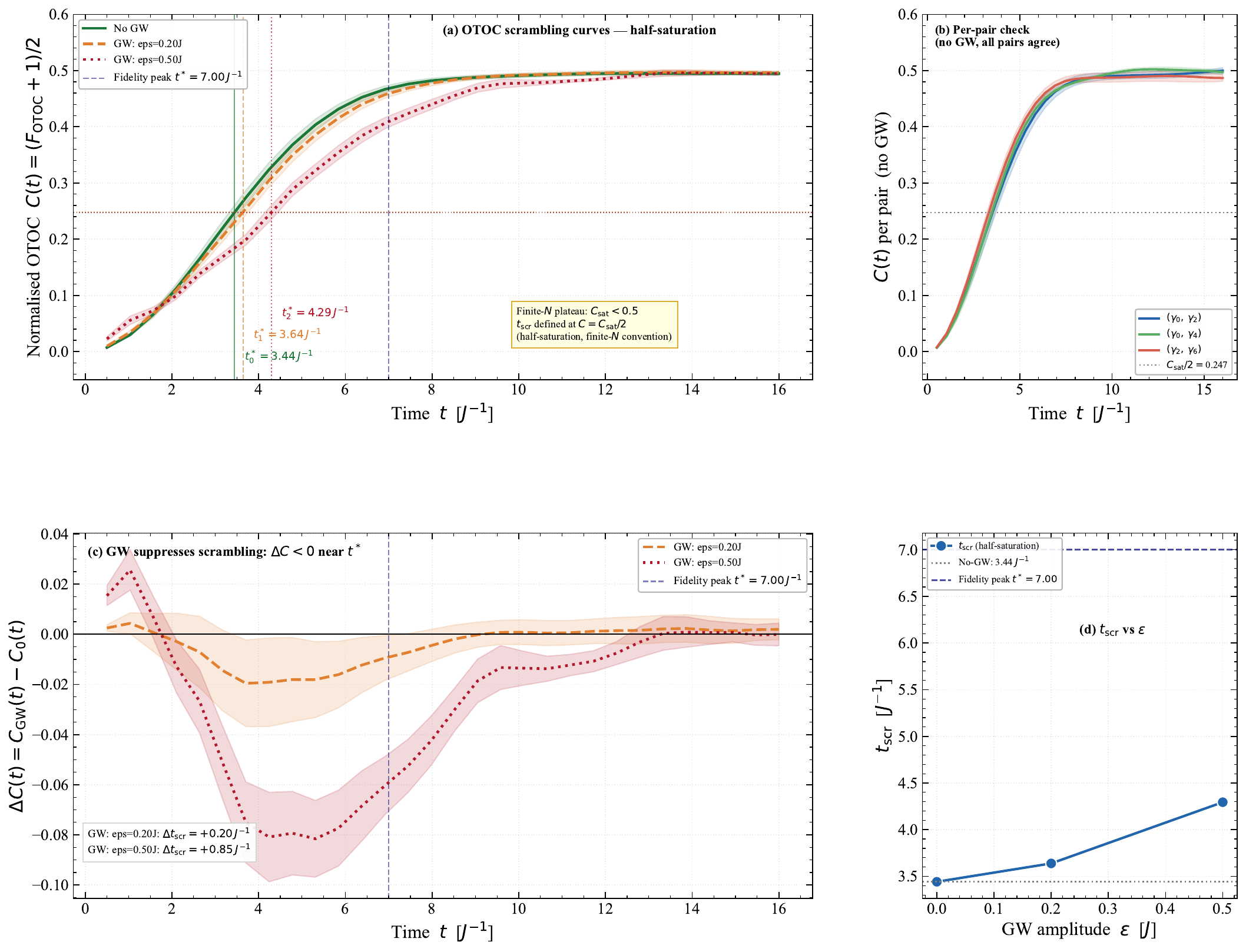}
    \caption{\label{fig:otoc}
OTOC scrambling diagnostic ($N=12$, $\beta J=2$, monochromatic drives at $\omega=1.5\,J$). (a)~Normalised OTOC $C(t)=(F_{\rm OTOC}+1)/2$ for no GW (green), $\varepsilon=0.20\,J$ (orange), and $0.50\,J$ (red). At finite $N=12$, $C(t)$ saturates at $C_{\rm sat}\approx 0.49 < 0.5$; the scrambling time $t_{\rm scr}$ is extracted via the half-saturation criterion $C(t_{\rm scr})=C_{\rm sat}/2$. The GW delays $t_{\rm scr}$ from $3.44\,J^{-1}$ (no GW) to $3.64\,J^{-1}$ ($\Delta t_{\rm scr}^{(\rm OTOC)}=+0.20\,J^{-1}$) and $4.29\,J^{-1}$ ($\Delta t_{\rm scr}^{(\rm OTOC)}=+0.85\,J^{-1}$), consistent in sign with the fidelity peak displacement of Fig.~\ref{fig:chirp}. (b)~$C(t)$ resolved by Majorana operator pair $(\gamma_0,\gamma_2)$, $(\gamma_0,\gamma_4)$, $(\gamma_2,\gamma_6)$ at $\varepsilon=0$; all pairs agree, confirming operator independence ($C_{\rm sat}/2=0.247$). (c)~OTOC deviation $\Delta C(t)=C_{\rm GW}(t)-C_0(t)$; $\Delta C<0$ near $ t^* = 7.00 \, J^{-1}$, returning to zero at late times, establishing a delay rather than a permanent disruption. (d)~Extracted $t_{\rm scr}$ vs.\ $\varepsilon$; monotone increase confirms that the GW-induced scrambling delay grows with drive amplitude.}
\end{figure*}

Section~\ref{sec:spectroscopy} characterized the channel response under monochromatic drives. We now turn to the inspiral-type chirp waveform of Eq.~(\ref{eq:chirp}), whose instantaneous frequency sweeps continuously across the thermal crossover during a single teleportation run. Three measurements are reported. We first show that the right boundary records the temporal structure of the chirp in a locally measurable strain response $S_R(t)$ [Fig.~(\ref{fig:detector})]. We then examine the effect of the same drive on the teleportation fidelity profile [Fig.~(\ref{fig:chirp})]. Finally, and independently of the teleportation protocol, we use an out-of-time-order correlator under a monochromatic drive to establish that the underlying many-body scrambling time is itself delayed [Fig.~(\ref{fig:otoc})]. \\

We begin with the boundary strain response, the locally measurable one-sided observable
\begin{equation}
  S_R(t)
  \;\equiv\;
  \bigl\langle H_{\rm strain}^R\bigr\rangle_t
  = \langle\psi(t)|\,H_{\rm strain}^R\,|\psi(t)\rangle.
  \label{eq:SR}
\end{equation}
We distinguish $S_R(t)$ carefully from the total boundary energy $\langle H_R\rangle_t$: it is the expectation value of the strain operator alone, which is the component of the response directly coupled to the GW field, not the full Hamiltonian expectation. Being a one-sided observable, $S_R(t)$ requires no reference ancilla and no two-boundary correlator \cite{schuster2022many}.\\
For this measurement the left preparation runs under the unperturbed $H_L$ only; the GW enters exclusively through $H_R(t) = H_R + \varepsilon_0 h(t) H_{\rm strain}^R$ during readout, with $\varepsilon_0 = 0.50\,J$. Within linear response, $S_R(t)$ satisfies the retarded convolution
\begin{equation}
  S_R(t)
  = \varepsilon_0\int_0^t
    \chi_R(t - t')\,h(t')\,dt',
  \label{eq:Kubo}
\end{equation}
where
\begin{equation}
  \chi_R(\tau)
  = -i\,\theta(\tau)\,
  \bigl\langle\bigl[H_{\rm strain}^R(\tau),\,
  H_{\rm strain}^R(0)\bigr]\bigr\rangle_{\rm TFD}
  \label{eq:chiret}
\end{equation}
is the retarded susceptibility of the right boundary in the post-coupling TFD state \cite{kubo1957statistical}.

Two features of Fig.~(\ref{fig:detector}) warrant physical discussion. First, $S_R(t)$ is a retarded rather than an instantaneous response. The drive envelope terminates at $t = t^* = 7.0\,J^{-1}$, yet $S_R$ remains non-zero well beyond that time, decaying from $0.044\,J$ at $t = t^*$ to the disorder noise floor only near $t \simeq 8.3\,J^{-1}$. The boundary therefore retains memory of the drive over a time of order the thermal correlation time $\beta = 2\,J^{-1}$. In the Fourier domain, Eq.~\eqref{eq:Kubo} gives $\tilde S_R(\omega) = \varepsilon_0\,\tilde\chi_R(\omega)\,\tilde h(\omega)$, and a delay of this size is what a retarded susceptibility with memory set by the thermal correlation time $\beta = 2\,J^{-1}$ produces. The measured lag is of that order, indicating that the boundary integrates the drive over roughly one thermal correlation time before responding, rather than following it instantaneously \cite{garcia2017analytical}. \\
Second, the envelope of $S_R(t)$ is not constant; it is largest during the early readout and attenuates at later times. A direct quantitative comparison with Fig.~(\ref{fig:spectroscopy}) is not warranted, since that figure reports the bilateral monochromatic response at $\varepsilon = 0.20\,J$ whereas $S_R(t)$ is measured under a right-boundary-only chirp at $\varepsilon_0 = 0.50\,J$. The observed attenuation is nonetheless qualitatively consistent with the same underlying thermal-crossover mechanism: the instantaneous frequency $\omega_{\rm inst}(t)$ of Eq.~(\ref{eq:omega_inst_def}) sweeps upward through $\omega_T$ and beyond during the readout window, and the retarded susceptibility $\tilde\chi_R(\omega)$ that controls $S_R$ decreases across that crossover. In the stationary-phase approximation, valid when $\dot\omega_{\rm inst} \ll \omega^{2}_{\rm inst}$, $\Sigma(\omega_{\rm inst}(t))$ is large at early readout times when $\omega_{\rm inst} \lesssim \omega_T$ and diminishes as $\omega_{\rm inst}$ rises, so that the time-dependent amplitude of $S_R(t)$ tracks the same low-pass behaviour in the time domain. \\
Quantitatively, the peak strain response $S_R^{\rm peak} = 0.072\,J$, attained at $t_{\rm peak} = 5.33\,J^{-1}$, at coupling scale $\varepsilon_0\,\|H_{\rm strain}^R\|_{\rm sp} = 2.5\,J$ gives the dimensionless susceptibility coefficient
\begin{equation}
  \bar\chi_R
  = \frac{S_R^{\rm peak}}
         {\varepsilon_0\,\|H_{\rm strain}^R\|_{\rm sp}\,h(t_{\rm peak})}
  = \frac{0.072\,J}{2.5\,J \times 0.681}
  = 0.043,
  \label{eq:chibar}
\end{equation}
where $h(t_{\rm peak}) = A(t_{\rm peak})\cos(\varphi(t_{\rm peak})) = 0.681$ is the full waveform value at the time of peak response, not the envelope maximum; the envelope $A(t_{\rm peak}) < 1$ and the cosine factor together reduce the instantaneous waveform value below unity at that moment. In our normalization, $\bar\chi_R$ serves as an empirical linear-response gain for the one-sided strain readout at $N=12$, $\beta J=2$. \\

We now turn to the effect of the same chirp on the teleportation fidelity. In the absence of a GW drive, the teleportation fidelity as a function of readout time $t_R$ rises to a maximum near $t_R \approx t^* = 7.00\, J^{-1}$ and then decreases. This peak marks the protocol-dependent decoding optimum: the readout time at which the left preparation, coupling, and right evolution conspire to maximize the overlap between the decoded state and the original message. Because the left and right boundaries share the same disorder realization and therefore have statistically identical scrambling scales on average, a perturbation that slows scrambling on the boundary dynamics can shift the decoding optimum to later $t_R$. \\
The chirp drive acts exclusively on the right boundary during the readout step, with peak amplitude $\varepsilon_0 = 0.50\, J$. The left-boundary preparation (backward and forward evolution) runs under the unperturbed $H_L$ throughout, because the chirp waveform $h(t) = 0$ for $t < 0$ by construction, and both preparation steps evolve at negative times. The GW therefore enters only through $H_R(t) = H_R + \varepsilon_0 h(t) H_{\rm strain}^R$ during the decoding step. The drive produces two observable changes in the fidelity profile (Fig.~\ref{fig:chirp}): (i)~the fidelity maximum shifts to later readout times, consistent with a drive-induced delay of the effective decoding dynamics; (ii)~the peak fidelity is reduced because the bilinear strain term perturbs the right-boundary readout evolution during Step~3, while the left preparation remains unperturbed as stated above. \\

The peak suppression $\Delta\mathcal{F}$ and peak displacement $\Delta t_{\rm scr}^{(\rm fid)}$ have distinct physical origins. The suppression arises from partial disruption of the right-boundary decoding dynamics during the GW-driven readout, analogous to the mechanism identified in Sec.~\ref{sec:amplitude} for the amplitude scan. The displacement reflects a change in the effective chaotic growth rate of the boundary: a bilinear perturbation $\varepsilon H_{\rm strain}$, being integrable in isolation, is expected to reduce the effective scrambling rate of the four-body SYK Hamiltonian \cite{garcia2018exact,lashkari2013towards}. \\
We state the status of this displacement carefully. The measured value is $\Delta t_{\rm scr}^{(\rm fid)} = +0.11\,J^{-1}$, obtained by three-point parabolic interpolation of a fidelity profile sampled on a readout grid of spacing $\delta t_R = 0.33\,J^{-1}$. The shift is therefore sub-grid, and propagating the disorder error bars through the interpolation gives an uncertainty comparable to the shift itself. We accordingly treat the fidelity-peak displacement as indicative of the direction of the effect and not as an independent quantitative measurement. The quantitative result is supplied by the OTOC diagnostic below, which is extracted from an independent observable and is resolved well above the disorder noise. \\
The scrambling-time shift is expected to be positive (a delay) and, to leading perturbative order, an even function of $\varepsilon$, as discussed in Appendix~\ref{app:scrambling}:
\begin{equation}
  \Delta t_{\rm scr}
  \;\sim\; c(\omega, \beta)\,\varepsilon^2
    + \mathcal{O}(\varepsilon^4),
  \label{eq:Dtscr}
\end{equation}
where $c(\omega, \beta)$ is a non-universal coefficient. We present this as a heuristic expectation based on the $\mathbb{Z}_2$ disorder symmetry argument \cite{kitaev2015simple,garcia2016spectral}: the positive sign $\Delta t_{\rm scr} > 0$ is our primary supported result, while the specific scaling with $\varepsilon$ is tested independently by the OTOC diagnostic below. \\
A key feature of Fig.~(\ref{fig:chirp}) is the late-time recovery: $\mathcal{F}(t_R)$ under the GW drive returns to the unperturbed baseline for $t_R \gg t^*$. This rules out permanent channel disruption and establishes the GW effect as a transient distortion of the scrambling dynamics. In holographic language, this is consistent with a transient deformation of the effective traversable channel rather than a permanent loss of transmissivity \cite{maldacena2018eternal,gao2021traversable}. \\
The fidelity-peak displacement measures the protocol-level decoding optimum and depends on the specific $(g^*, t^*)$ operating point. To establish that the effect reflects a change in the intrinsic scrambling dynamics of the SYK Hamiltonian, which is independent of the protocol, we compute the out-of-time-order correlator (OTOC) \cite{LarkinOvchinnikov1969,kitaev2015simple} directly on the left-boundary TFD state, without the message qubit or coupling step \cite{xu2024scrambling}. For two Majorana operators $W = \gamma_i$ and $V = \gamma_j$ on the left boundary ($i \neq j$), the thermal OTOC is
\begin{equation}
  F_{\rm OTOC}(t) =
  \mathrm{Re}\bigl[
    \langle\mathrm{TFD}|
    W(t)\,V\,W(t)\,V
    |\mathrm{TFD}\rangle
  \bigr],
  \label{eq:FOTOC}
\end{equation}
where $W(t) = U_L^\dagger(t)\,\gamma_i\, U_L(t)$ is the Heisenberg-picture operator under the left SYK Hamiltonian (with GW drive where applicable). For the GW-driven cases, the Heisenberg evolution $U_L(t)$  includes the bilinear strain deformation $H_L + \varepsilon h(t) H_{\rm strain}^L$ on the left boundary, so that $t_{\rm scr}$ measures the GW effect on left-boundary scrambling directly. The monochromatic drive ($\omega = 1.5\, J$) is applied to both boundaries for the OTOC measurement, distinct from the right-boundary-only chirp of Fig.~(\ref{fig:chirp}); the two experiments therefore probe the same physical effect through different protocols.
The initial value is fixed by the Majorana algebra: for $i \neq j$,
\begin{equation}
  \gamma_i\gamma_j\gamma_i\gamma_j
  = -(\gamma_i)^2(\gamma_j)^2
  = -\mathbb{I} \qquad (i \neq j),
  \label{eq:Majorana_init}
\end{equation}
giving $F_{\rm OTOC}(0) = -1$. The normalized diagnostic
\begin{equation}
  C(t) = \frac{F_{\rm OTOC}(t) + 1}{2}
  \label{eq:Ct}
\end{equation}
therefore satisfies $C(0) = 0$ and grows toward $C_{\rm sat}$ as the system scrambles. The growth of $C(t)$ directly tracks operator spreading: as $W(t)$ evolves from a simple local Majorana into a complex many-body operator, $F_{\rm OTOC}(t)$ grows from $-1$ toward $0$, and $C(t)$ grows from $0$ toward $C_{\rm sat}$. \\
At large $N$, $C(t)$ saturates at $C_{\rm sat} = 1/2$ in the fully scrambled state. At finite $N$, the plateau falls below this value due to finite-size and operator-normalisation effects specific to the TFD state and the Majorana operator pair chosen \cite{hosur2016chaos,roberts2017chaos}. Rather than imposing the large-$N$ limit, we estimate $C_{\rm sat}$ directly from the measured late-time plateau in each dataset, and define the scrambling time via the half-saturation criterion
\begin{equation}
  C(t_{\rm scr}) = \frac{C_{\rm sat}}{2}.
  \label{eq:half_sat}
\end{equation}
For $N=12$ we find a late-time plateau $C_{\rm sat}\approx 0.49$ (for the operator pairs considered), giving the working threshold $C_{\rm sat}/2\approx 0.247$. This criterion is robust at finite $N$ because it is defined relative to the empirically measured plateau rather than imposing the large-$N$ value $1/2$. Scrambling times are extracted by linear interpolation between adjacent time points where $C(t)$ crosses the threshold \cite{garcia2018exact}. \\

Fig.~(\ref{fig:otoc}) Panel (a) shows $C(t)$ for the three cases: no GW, $\varepsilon = 0.20\,J$, and $\varepsilon = 0.50\,J$, using a monochromatic drive at $\omega = 1.5\,J$ (averaged over three Majorana pairs and $N_{\rm avg}=5$ disorder realisations). All three curves plateau at $C_{\rm sat} \approx 0.49$, and the GW-driven curves are shifted consistently to later times. The extracted scrambling times are
\begin{equation}
  t_{\rm scr}^{(0)} = 3.44\,J^{-1},\quad
  t_{\rm scr}^{(0.20)} = 3.64\,J^{-1},\quad
  t_{\rm scr}^{(0.50)} = 4.29\,J^{-1},
  \label{eq:tscr_results}
\end{equation}
giving delays $\Delta t_{\rm scr}^{(\rm OTOC)} = +0.20\,J^{-1}$ and $+0.85\,J^{-1}$ at the two GW amplitudes. Both delays are positive: the GW retards scrambling. Panel~(b) shows that three independent Majorana operator pairs give identical $C(t)$ at $\varepsilon = 0$, confirming that $t_{\rm scr}$ is a property of the many-body dynamics rather than the choice of probe operator. Panel~(c) shows the OTOC deviation $\Delta C(t) = C_{\rm GW}(t) - C_0(t)$: it is negative near $t^* = 7.00\, J^{-1}$; the GW-driven system has scrambled less information by this time than the unperturbed system and returns to zero at late times, independently confirming the transient character of the GW effect. Panel~(d) shows $t_{\rm scr}$ as a monotone function of $\varepsilon$, with the delay growing from $+0.20\,J^{-1}$ at $\varepsilon = 0.20\,J$ to $+0.85\,J^{-1}$ at $\varepsilon = 0.50\,J$. The growth is superlinear in $\varepsilon$, consistent with a perturbative even-in-$\varepsilon$ correction as expected from the $\mathbb{Z}_2$ disorder symmetry, though the two data points are insufficient to determine the exponent precisely. \\
Both delays are resolved well above the disorder noise. A bootstrap analysis over the $N_{\rm avg}=5$ disorder realizations, jointly resampling the driven and unperturbed OTOC traces so as to preserve their correlation, yields
\begin{equation}
\begin{split}
  \Delta t_{\text{scr}}(0.20\,J) &= +0.20 \pm 0.014\,J^{-1}, \\
  \Delta t_{\text{scr}}(0.50\,J) &= +0.85 \pm 0.034\,J^{-1},
\end{split}
\label{eq:tscr_significance}
\end{equation}
so that each delay exceeds its bootstrap standard error by more than an order of magnitude, and no resampled realization of either ensemble produces a negative delay. We deliberately refrain from converting these ratios into Gaussian tail probabilities: with five disorder realizations the appropriate reference distribution carries four degrees of freedom, and a nonparametric bootstrap over five samples cannot resolve tail probabilities below $\sim\!10^{-2}$. The quoted uncertainties should accordingly be read as noise-floor estimates rather than as calibrated significance levels. \\
Taking the two amplitudes together gives a two-point power-law exponent $n = \ln(4.25)/\ln(2.5) \approx 1.6$, to be compared with the even-power expectation $n=2$ discussed in Appendix~\ref{app:schwarzian}. Two points cannot fix an exponent, and at $\beta J = 2$ and $N=12$ the system lies outside the regime in which the leading-order Schwarzian estimate applies. We therefore regard the data as consistent with an even-power response in $\varepsilon$ without claiming quantitative agreement with $\varepsilon^2$. \\
The fidelity peak displacement and the OTOC scrambling delay are distinct observables that report on the same underlying effect through different windows. Their connection follows from the Gao-Jafferis-Wall construction \cite{gao2017traversable}: in the WITP, the teleportation fidelity at readout time $t_R$ is related to the two-point function of the boundary at $t_R$ in the perturbed TFD state. The fidelity peak occurs at the $t_R$ that maximizes the decoded overlap, which is correlated with the boundary scrambling time $t_{\rm scr}^{(R)}$, which is the time at which information injected at the left boundary has traversed the wormhole and can be recovered on the right \cite{yoshida2017efficient,schuster2022many}. A drive that increases a characteristic scrambling/decoding timescale on the right boundary is expected to shift the calibrated fidelity peak in the same direction, motivating the heuristic relation:
\begin{equation}
  \Delta t_{\rm scr}^{(\rm fid)}
  \;\approx\; \Delta t_{\rm scr}^{(R)}.
  \label{eq:fid_scr_relation}
\end{equation}
The OTOC diagnostic measures $t_{\rm scr}$ independently, without the protocol, providing a direct check on this interpretation. The magnitudes differ: $\Delta t_{\rm scr}^{(\rm fid)} = +0.11\,J^{-1}$ (chirp, $\varepsilon_0 = 0.50\,J$) versus $\Delta t_{\rm scr}^{(\rm OTOC)} = +0.20\,J^{-1}$ (monochromatic, $\varepsilon = 0.20\,J$). We do not attempt a quantitative comparison of these two numbers, and we note explicitly that a naive amplitude-based estimate does not account for the difference. The chirp waveform $h(t)=A(t)\cos\varphi(t)$ has root-mean-square amplitude $\varepsilon_{\rm rms} = \varepsilon_0\langle A^2\cos^2\varphi\rangle_t^{1/2} = 0.22\,J$ when averaged over the full readout window, and $0.25\,J$ over $[0,t^*]$, which \emph{exceeds} the OTOC reference amplitude $0.20\,J$; an even-power response in $\varepsilon$ alone would therefore predict a larger fidelity-peak displacement than the OTOC delay, whereas the opposite is observed. The two measurements differ in three further respects that no amplitude comparison captures. The chirp acts only on the right boundary and only during the readout step, whereas the OTOC is measured on the left boundary under a drive applied throughout the evolution. The chirp sweeps its instantaneous frequency from $\omega_T$ up to $5.5\,J$, so only a fraction of its spectral weight lies in the quasi-static region of maximal response. And the fidelity peak and the OTOC half-saturation time are distinct observables with different sensitivities. What we claim from this comparison is the agreement in sign, not in magnitude. \\
This sign agreement rules out protocol mismatch or numerical artifact as the origin of the fidelity distortion, and directly supports the interpretation of the WITP as a device sensitive to GW-induced changes in the chaotic scrambling dynamics of the SYK boundary. \\
We also note a striking holographic parallel. In rigorous AdS/CFT shockwave calculations, a neutral infalling shell scrambles information according to the standard Lyapunov exponent, producing no positive time delay. However, when the infalling shell carries $U(1)$ charge or angular momentum, it encounters a repulsive potential barrier near the inner Cauchy horizon of the Reissner-Nordström or Kerr black hole and bounces, physically delaying the collision between the shockwave and the infalling Hayden-Preskill signal \cite{horowitz2022bouncing}. \\
On the asymptotic boundary, this geometric bounce manifests precisely as a positive shift $\Delta t_{\rm scr} > 0$, rather than a suppression of $\lambda_L$ \cite{prihadi2025scrambling}. The empirical $\Delta t_{\rm scr} > 0$ found here with both the fidelity peak displacement and the OTOC diagnostic agreeing in sign is qualitatively consistent with the phenomenological Floquet drive acting as an effective ``charge-like'' injection that modifies the interior bulk geometry of the dual black hole, pushing the information collision deeper into the bulk. We stress that this parallel is purely phenomenological: the present SYK model carries no $U(1)$ charge or angular momentum, and a precise holographic identification of the Floquet drive with a charged or rotating bulk perturbation would require a Schwarzian effective-theory analysis, which we reserve for future work.

\section{Genuine effects vs Protocol Mismatch} \label{sec:reopt}
\begin{figure*}
    \centering
    \includegraphics[width=\linewidth]{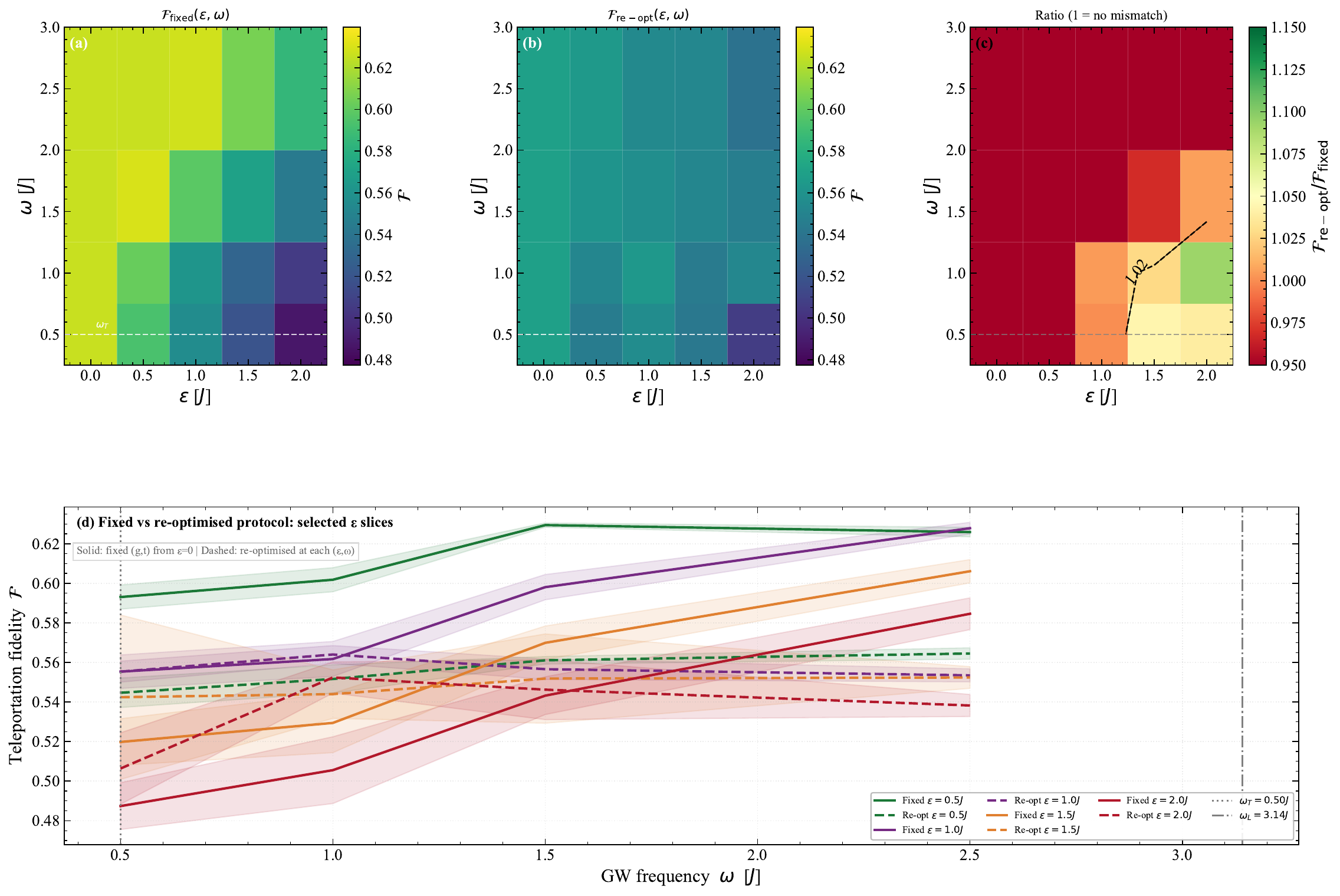}
    \caption{\label{fig:reopt}
Protocol mismatch analysis ($N=12$, $\beta J=2$, $N_{\rm avg}=5$).(a,b)~Heatmaps of $\mathcal{F}_{\rm fixed}(\varepsilon,\omega)$ and $\mathcal{F}_{\rm re\text{-}opt}(\varepsilon,\omega)$ over the GW parameter grid; $(g^*,t^*)$ are fixed at their $\varepsilon=0$ optima in (a) and independently re-optimised at each $(\varepsilon,\omega)$ point in (b). (c)~Ratio $\mathcal{F}_{\rm re\text{-}opt}/ \mathcal{F}_{\rm fixed}$ (unity indicates no benefit from retuning). The ratio is consistent with 1 in the sensing regime ($\varepsilon\lesssim J$), confirming genuine GW-induced degradation; it rises to $\sim\!1.15$ only in the strong slow-drive corner ($\varepsilon\gtrsim 1.5\, J$, $\omega\lesssim\omega_T$). (d) Selected $\varepsilon$-slices ($\varepsilon = 0.5\, J$ and $1.0\, J$) of $\mathcal{F}$ vs.\ $\omega$, with fixed (solid) and re-optimized (dashed) protocols; the two are indistinguishable within statistical uncertainty throughout the sensing regime. The thermal ($\omega_T=0.50\,J$) and MSS chaos scale ($\omega_L=3.14\,J$) scales are marked.}
\end{figure*}

A central interpretive question for deploying any driven quantum channel is whether the observed fidelity degradation reflects a genuine dynamical effect of the drive, or whether it arises because the protocol was calibrated for the unperturbed Hamiltonian and is therefore suboptimal under driving \cite{almheiri2020replica}. We distinguish these two contributions through an explicit re-optimisation test Fig.~(\ref{fig:reopt}): we compare the fidelity obtained with the fixed $\varepsilon = 0$ protocol parameters $(g^*, t^*)$ against the fidelity obtained by independently optimising $(g, t)$ at each $(\varepsilon, \omega)$ grid point under the live drive. \\
The decomposition is algebraically exact by construction; in practice, since the re-optimization uses a coarse grid over three disorder seeds and the ratio is evaluated over $N_{\rm avg} = 5$ seeds, the mismatch term carries a statistical uncertainty of order $\sigma_{\rm dis}/\sqrt{N_{\rm avg}}$ and individual grid points may show $r$ slightly below unity due to disorder noise, not genuine protocol degradation. Denote by $\mathcal{F}_{\rm fixed}(\varepsilon, \omega)$ the fidelity evaluated with the $\varepsilon = 0$ optima $(g^*_0, t^*_0)$, and by $\mathcal{F}_{\rm re\text{-}opt}(\varepsilon, \omega)$ the fidelity maximised over $(g, t)$ at the given drive conditions:
\begin{equation}
  \mathcal{F}_{\rm re\text{-}opt}(\varepsilon, \omega)
  = \max_{g,\,t}\,
  \mathcal{F}(g, t;\, \varepsilon, \omega).
  \label{eq:Freopt}
\end{equation}
The total fidelity loss from the unperturbed baseline
$\mathcal{F}_0$ decomposes as
\begin{equation}
  \underbrace{\mathcal{F}_0
    - \mathcal{F}_{\rm fixed}}_{\text{total loss}}
  =
  \underbrace{\mathcal{F}_0
    - \mathcal{F}_{\rm re\text{-}opt}}_{\text{genuine}}
  +
  \underbrace{\mathcal{F}_{\rm re\text{-}opt}-\mathcal{F}_{\rm fixed}}_{\text{mismatch (definition)}}.
  \label{eq:decomp}
\end{equation}
The first term on the right is the genuine GW-induced channel degradation: the fidelity loss that persists even when the protocol is optimally recalibrated for the drive. The second term is the protocol mismatch: additional loss that would be recovered by recalibrating $(g, t)$. A clean claim of drive-induced degradation requires that the genuine component be nonzero; the mismatch component is an experimental artifact that vanishes with a sufficiently agile calibration procedure. \\
The key diagnostic is the ratio
\begin{equation}
  r(\varepsilon, \omega)
  = \frac{\mathcal{F}_{\rm re\text{-}opt}(\varepsilon, \omega)}
         {\mathcal{F}_{\rm fixed}(\varepsilon, \omega)}
  = 1 + \frac{\mathcal{F}_{\rm re\text{-}opt}
              - \mathcal{F}_{\rm fixed}}
             {\mathcal{F}_{\rm fixed}}.
  \label{eq:ratio}
\end{equation}
$r = 1$ everywhere means all degradation is genuine and no fidelity can be recovered by recalibrating; $r > 1$ means a substantial fraction of the loss is a calibration artifact \cite{bandyopadhyay2023universal}. We map $r(\varepsilon, \omega)$ over the grid $\varepsilon \in \{0, 0.5, 1.0, 1.5, 2.0\}\,J$ and $\omega \in \{0.5, 1.0, 1.5, 2.5\}\,J$ by grid-searching over $(g, t) \in [5, 35] \times [3, 14]\,J^{-1}$ using three disorder seeds for the re-optimisation and $N_{\rm avg} = 5$ for the final disorder average. The reduced $N_{\rm avg}$ relative to the main results ($N_{\rm avg} = 20$) reflects the computational cost of the double loop, but is sufficient to determine the sign and approximate magnitude of $r-1$ at each grid point. \\

The results establish two distinct parameter regions. Throughout the sensing regime $\varepsilon \lesssim J$, the ratio map is flat at $r \approx 1$ across all four $\omega$ values in the grid. Fig.~(\ref{fig:reopt} Panel~(d)) shows the $\varepsilon = 0.5\,J$ and $1.0\,J$ slices explicitly: the fixed-protocol and re-optimized-protocol curves are indistinguishable within the statistical uncertainty of $N_{\rm avg} = 5$, showing that the re-optimized and fixed protocols are indistinguishable in the sensing regime, consistent with predominantly genuine drive-induced degradation. This is the parameter region used for all quantitative results in Secs.~\ref{sec:amplitude}--\ref{sec:chirp} (where $\varepsilon = 0.20\,J$), and the result directly validates the interpretation of those fidelity signals as genuine GW-induced channel degradation.\\
A qualitatively different behavior appears in the strong slow-drive corner, $\varepsilon \gtrsim 1.5\, J$ combined with $\omega \lesssim \omega_T = 0.50\,J$. Here $r$ reaches up to $\sim\!1.15$, i.e. retuning yields up to a $\sim\!15\%$ relative gain in fidelity compared to the fixed protocol in this corner of parameter space. The physical origin is clear from the quasi-static limit: at $\omega \ll \omega_T$ the drive appears static on the thermal timescale, so the boundary evolves under an effective Hamiltonian
\begin{equation}
  H_{\rm eff} \approx H_\alpha + \varepsilon\,H_{\rm strain}^\alpha,
  \label{eq:Heff_static}
\end{equation}
which has a different spectrum and scrambling dynamics from those of $H_\alpha$ alone \cite{almheiri2020replica}. The optimal coupling $g^*$ and readout time $t^*$ for the traversable wormhole are set by the scrambling structure of $H_{\rm eff}$~, and both shift once the quasi-static deformation becomes large enough to appreciably modify the effective dynamics over the protocol window. Crucially, even in this corner, $r \lesssim 1.15$: the protocol mismatch accounts for at most $15\%$ of the total fidelity loss at $(\varepsilon, \omega) = (2.0\,J, 0.5\,J)$, the remaining degradation persists even after retuning, indicating a genuine drive-induced change in the effective channel dynamics. \\
The interpretation is therefore unambiguous. The fidelity degradation reported in Secs.~\ref{sec:amplitude}--\ref{sec:chirp} is dominated by genuine GW-induced effects on the SYK scrambling dynamics, not by a stale protocol calibration. In the sensing regime $\varepsilon \lesssim J$, which is the operationally relevant weak-drive regime, the mismatch contribution is negligible by construction, and the fidelity signal is a direct measure of the GW perturbation. \\

\section{System-size scaling}\label{sec:scaling}
All quantitative results presented in Secs.~\ref{sec:amplitude}--\ref{sec:reopt} were obtained at $N = 12$. A necessary condition for the GW-induced effects to be physically meaningful rather than artifacts of the small Hilbert space $d = 2^{N/2} = 64$ is that they persist and do not diminish as $N$ increases toward the holographic large-$N$ limit. We test this by repeating the fidelity measurements at $N = 10$, $12$, $14$ and $16$ Majorana modes, keeping $\beta J = 2$ and $N_{\rm avg} = 50$ fixed throughout.\\

\begin{figure*}
    \centering
    \includegraphics[width=\linewidth]{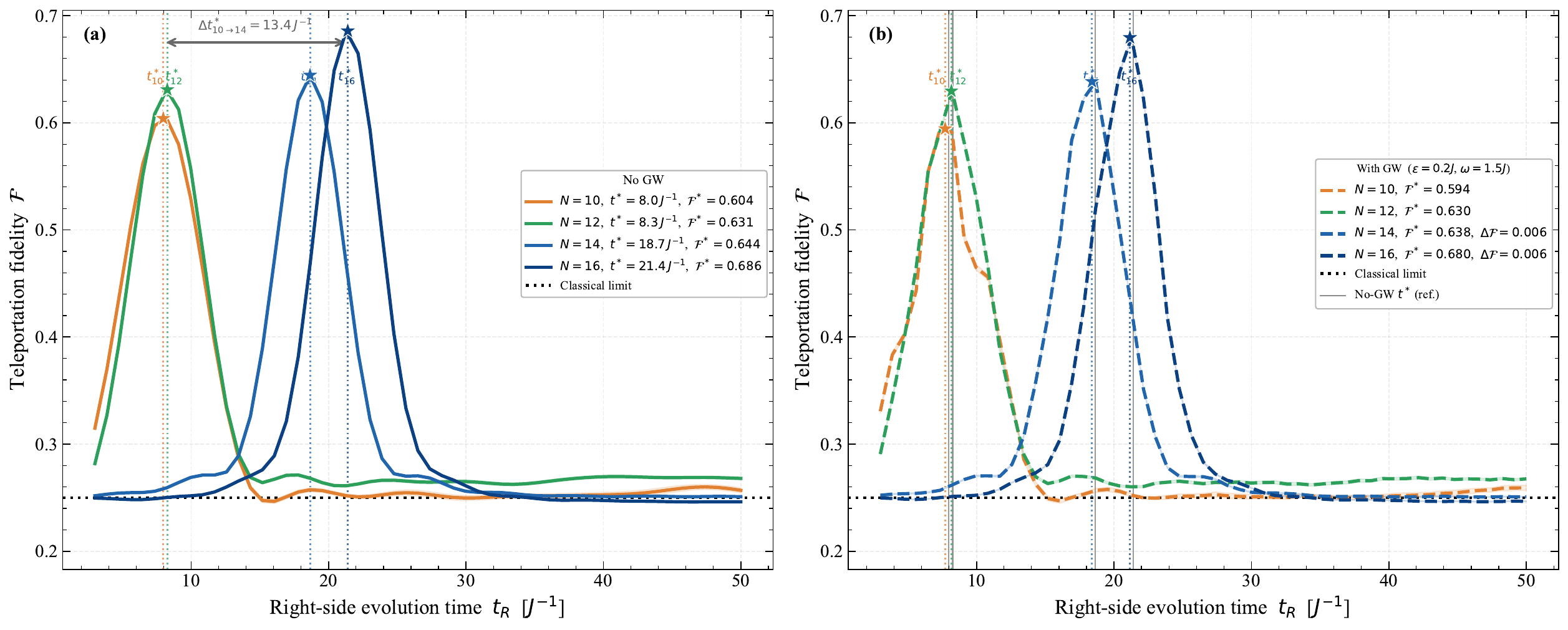}
\caption{\label{fig:scaling}
System-size scaling at $\beta J=2$, $N_{\rm avg}=50$. (a)~Fidelity profiles $\mathcal{F}(t_R)$ without GW for $N=10$ (peak at $t^*=7.97\,J^{-1}$, $\mathcal{F}^*=0.604$), $N=12$ ($t^*=8.27\,J^{-1}$, $\mathcal{F}^*=0.6309$), $N=14$ ($t^*=18.65\,J^{-1}$, $\mathcal{F}^*=0.6444$) and $N=16$ ($t^*=21.40 \,J^{-1}$,$\mathcal{F}^*=0.6858$). The span of optimal teleportation times across system sizes, $\Delta t^*_{10\to 16}=13.4\,J^{-1}$ (annotated arrow), grows monotonically with $N$. (b)~Fidelity profiles under the GW drive $(\varepsilon,\omega)=(0.20\,J,\,1.5\,J)$ for the same $N$ values ($\mathcal{F}^*=0.5944$, $0.6296$, $0.6384$, $0.6795$); dashed vertical lines mark the no-GW reference peaks from panel (a). The GW-induced fidelity suppression is non-zero across all four system sizes and shows no systematic decrease with $N$, supporting its interpretation as a genuine many-body effect rather than a finite-size artifact.}
    
\end{figure*}

The two panels of Fig.~(\ref{fig:scaling}) isolate two complementary aspects of the $N$-dependence: the unperturbed channel properties (panel a) and the GW response at fixed
$(\varepsilon, \omega) = (0.20\,J, 1.5\,J)$ (panel b). Panel~(a) establishes the $N$-dependence of the unperturbed channel. The baseline fidelity $\mathcal{F}^*$ increases with $N$,
\begin{equation}
\begin{split}
  \mathcal{F}^*_{(N=10)} &= 0.6040, \qquad
  \mathcal{F}^*_{(N=12)}  = 0.6309, \\
  \mathcal{F}^*_{(N=14)} &= 0.6444, \qquad
  \mathcal{F}^*_{(N=16)}  = 0.6858,
\end{split}
\label{eq:F0_scaling}
\end{equation}
with increments that suggest an approach toward a finite large-$N$ value, consistent with the expectation that at fixed $\beta J$ the protocol fidelity approaches a nonzero limit as $N$ grows. The $N = 12$ entry $\mathcal{F}^{*}_{(N=12)} = 0.6309$ reflects an independent re-optimisation of $(g^{*}, t^{*})$ for each $N$ in this scaling comparison, yielding $t^{*} = 8.27\,J^{-1}$ at $N = 12$. The main-text baseline $\mathcal{F}_0 = 0.626$ corresponds to the fixed operating point $(g^{*} = 12.0\,J,\; t^{*} = 7.0\,J^{-1})$ used throughout Secs.~\ref{sec:amplitude}--\ref{sec:reopt}; the difference from $\mathcal{F}^*_{(N=12)} = 0.6309$ is within the disorder noise for $N_{\rm avg} = 20$. In holographic language, increasing $N$ at fixed $\beta J$ moves the finite-size numerics toward the regime where the low-energy effective description becomes more accurate. The observed monotone increase of $\mathcal{F}^*$ is consistent with finite-$N$ corrections diminishing as the Hilbert space grows. \\
The optimal readout time $t^*$ also increases monotonically with $N$:
\begin{equation}
\begin{split}
  t^*_{(N=10)} &= 7.97\,J^{-1}, \qquad t^*_{(N=12)} = 8.27\,J^{-1}, \\
  t^*_{(N=14)} &= 18.65\,J^{-1}, \qquad t^*_{(N=16)} = 21.40\,J^{-1},
\end{split}
\label{eq:tstar_scaling}
\end{equation}
spanning $\Delta t^{*}_{10\to 16} = 13.4\,J^{-1}$. The growth is monotone but irregular: (increments quoted from the unrounded interpolated peak positions) the $N=12\to14$ increment ($10.39\,J^{-1}$) is far larger than the $N=10\to12$ increment ($0.29\,J^{-1}$), while the $N=14\to16$ increment ($2.75\,J^{-1}$) is intermediate, consistent with the decoding optimum settling as the dimension $d$ grows beyond $128$. \\
This non-smooth behaviour reflects the rapid growth of the Hilbert space dimension $d = 2^{N/2}$ (from $32$ to $64$ to $128$ to $256$ as $N$ steps from$10$ to $12$ to $14$ to $16$): as $d$ doubles, the discrete many-body spectrum reorganizes and the finite-size structure of the protocol fidelity profile can shift substantially, pushing the decoding optimum to later $t_R$ \cite{kobrin2021many}. The fidelity-peak location is additionally sensitive to near-degeneracies in the finite spectrum that reorganize each time $d$ doubles \cite{cotler2017black}.\\
We emphasize that the optimal readout time $t^{*}$ is not the same quantity as the Schwarzian scrambling time $t_{\rm scr} \sim (\beta/2\pi)\ln N$ \cite{kitaev2015simple,maldacena2016bound,garcia2016spectral,sunderhauf2019quantum}. The latter changes only logarithmically with $N$ (at fixed $\beta$). The decoding optimum $t^{*}$ depends on the full protocol structure, including the forward and backward left evolution and the coupling step, and is sensitive to the discrete spectral organization of the finite-$N$ system in a way that $t_{\rm scr}$ is not. The large $N=12\to 14$ jump, therefore, reflects a qualitative reorganization of the many-body spectrum as $d$ doubles from $64$ to $128$, rather than a violation of the expected $\ln N$ growth of the scrambling time. \\
To confirm that $t^{*} = 18.65\,J^{-1}$ is a genuine feature rather than a finite-sample outlier, we verified that the disorder-averaged fidelity profile retains its maximum at this location over the full $N_{\rm avg} = 50$ ensemble, with a disorder standard error of $3\times10^{-4}$ on the peak height, consistent with a physical feature of the $N=14$ spectrum. \\
Fig.~(\ref{fig:scaling}) Panel~(b) tests whether the GW signal inherits this $N$-dependence. The drive-induced fidelity suppression is defined as the difference between the respective peak values, $\Delta\mathcal{F}^*(N) \equiv \mathcal{F}^*_{\rm no\,GW}(N) - \mathcal{F}^*_{\rm GW}(N)$, where each peak is located independently by scanning $t_R$. This peak-to-peak definition compares the maximum attainable fidelity under each condition and is the quantity computed in our numerics. The statistical uncertainty on this difference is
\begin{equation}
  \sigma(\Delta\mathcal{F}^*)
  = \sqrt{\,\sigma_{\rm cl}^2(t^*_N) + \sigma_{\rm gw}^2(t^*_N)\,},
  \label{eq:sigma_def}
\end{equation}
where $\sigma_{\rm cl}(t^*_N) = \sigma_{\rm dis}^{\rm cl}/\sqrt{N_{\rm avg}}$ and $\sigma_{\rm gw}(t^*_N) = \sigma_{\rm dis}^{\rm gw}/\sqrt{N_{\rm avg}}$ are the disorder standard errors of the no-GW and GW fidelity profiles evaluated at the peak time $t^*_N$, computed from $N_{\rm avg}=50$ independent disorder realisations at each $N$. The two peak times coincide to within the grid spacing at $N=12,14,16$; at $N=10$ they differ by one grid step, but $\sigma_{\rm gw}$ is unchanged between the two locations, so the quoted uncertainties are unaffected. The measured suppressions and their standard errors are
\begin{equation}
\begin{split}
  \Delta\mathcal{F}^*_{(N=10)} &= 0.0095
    \quad [\sigma(\Delta\mathcal{F}^*)=0.00167], \\
  \Delta\mathcal{F}^*_{(N=12)} &= 0.0012
    \quad [\sigma(\Delta\mathcal{F}^*)=0.00120], \\
  \Delta\mathcal{F}^*_{(N=14)} &= 0.0059
    \quad [\sigma(\Delta\mathcal{F}^*)=0.00051], \\
  \Delta\mathcal{F}^*_{(N=16)} &= 0.0063
    \quad [\sigma(\Delta\mathcal{F}^*)=0.00037].
\end{split}
\label{eq:DF_significance}
\end{equation}
The standard error $\sigma(\Delta\mathcal{F}^*)$ falls by roughly a factor of four between $N=10$ and $N=16$, reflecting the improved self-averaging of the many-body fidelity as the Hilbert-space dimension grows from $d=32$ to $d=256$: at larger $d$ the density of many-body states increases and individual disorder realizations produce more consistent fidelity peaks, reducing $\sigma(\Delta\mathcal{F}^*)$ even at fixed $N_{\rm avg}=50$. Expressed as ratios of the measured suppression to its standard error, the four system sizes give $5.7$, $1.0$, $11.6$, and $17.0$. Three of the four, therefore, establish a non-zero drive-induced suppression at a level that cannot plausibly be ascribed to disorder noise. We quote these as ratios rather than as tail probabilities: the four measurements share the same drive parameters, protocol, and numerical implementation, so they are not independent statistical tests, and combining them multiplicatively would not be justified. The $N=12$ measurement is not resolved above the disorder noise at this ensemble size; with four system sizes and a single drive frequency, we are unable to identify the reason, and we discuss this further in Appendix~\ref{app:finiteN}. \\
The improvement in significance is driven principally by the fall in $\sigma(\Delta\mathcal{F}^*)$ rather than by growth of $\Delta\mathcal{F}^*$ itself, which is not monotone in $N$. Because $\Delta\mathcal{F}^*$ compares the two profile maxima, it is by construction insensitive to any displacement of the decoding optimum and measures only the reduction in maximum attainable fidelity, that is, direct degradation of the channel; this is consistent with the re-optimisation analysis of Sec.~\ref{sec:reopt}. The fidelity peak does nonetheless sharpen with $N$: the curvature $|\mathcal{F}''(t^*)|$ measured from the profiles of panel~(a) increases monotonically from $0.041\,J^{2}$ at $N=10$ to $0.076\,J^{2}$ at $N=16$, so a given displacement of the optimum costs progressively more fidelity at a fixed calibrated readout time. That mechanism bears on the fixed-calibration measurements of Secs.~\ref{sec:amplitude}--\ref{sec:spectroscopy} rather than on $\Delta\mathcal{F}^*$, and is distinct from the raw susceptibility $\Sigma(\omega)$ growing: with $\|H_{\rm strain}^{\alpha}\|_{\rm sp} = 5J$ fixed at all $N$, $\Sigma$ approaches a finite $N$-independent limit. \\
The most important conclusion from panel~(b) is the absence of any suppression or sign change in $\Delta\mathcal{F}^*$ as $N$ increases. A finite-size artifact such as an accidental resonance between the drive frequency $\omega = 1.5\, J$ and a few-body level spacing particular to a small-$d$ spectrum would not be expected to persist as $N$ increases, since the many-body spectrum and its level spacings reorganize substantially as the Hilbert space grows. The observed behavior is the opposite: the effect is nonzero and nonvanishing across all four system sizes, with no systematic suppression as $N$ increases from $10$ to $16$, supporting its interpretation as a genuine feature of the many-body scrambling dynamics rather than a finite-size coincidence. We note that the scaling run does not resolve any shift of the fidelity peak itself. The readout grid used here has a spacing of $0.87\,J^{-1}$, and the interpolated peak displacements across all four system sizes are less than a third of that spacing. The scaling comparison, therefore, constrains the drive-induced change in peak \emph{height}, $\Delta\mathcal{F}^*$, and not the peak position; the scrambling delay itself is established by the OTOC diagnostic of Sec.~\ref{sec:chirp}. \\
Taken together, the scaling results of Fig.~(\ref{fig:scaling}) show that the WITP GW response is not suppressed over the accessible range $N \in \{10, 12, 14, 16\}$. The unperturbed channel quality ($\mathcal{F}^*$), the protocol decoding optimum ($t^*$), and the GW sensitivity ($\Delta\mathcal{F}^*$) are all non-zero and show no systematic suppression with $N$, consistent with their large-$N$ limits being non-zero and finite \cite{gao2017traversable,maldacena2016bound}. A quantitative extrapolation to large $N$ requires extending the simulation to $N \gg 16$, which is computationally inaccessible by exact diagonalization but may be tractable by tensor-network or large-$N$ saddle-point methods \cite{cotler2017black,garcia2016spectral}; we leave this to future work.
\section{Summary and Conclusion}

We have studied how a GW-inspired bilinear boundary drive perturbs the wormhole-teleportation fidelity of an $N$-Majorana SYK model prepared in a thermofield-double state at $\beta J = 2$. The central finding is that the drive produces a coherent, frequency-dependent (low-pass), amplitude-dependent suppression of the teleportation channel, traceable through an independent OTOC diagnostic to a delay in the many-body scrambling time.\\
The amplitude scan at fixed frequency establishes the perturbative structure of the response. The fidelity loss $\Delta\mathcal{F}(\varepsilon)$ vanishes at $\varepsilon = 0$ and grows as an even power series in $\varepsilon$, consistent with the $\mathbb{Z}_2$ symmetry of the SYK coupling distribution, which forbids odd-order contributions. The channel remains above the classical random-guess baseline, $\mathcal{F} = 1/4$, across the full amplitude range studied, and the suppression accelerates in the strong-drive regime, where the bilinear strain perturbation becomes comparable to the typical spectral scale of the unperturbed dynamics.\\
The frequency scan reveals that the channel is most sensitive in the quasi-static regime $\omega \lesssim \omega_T = \beta^{-1}$, where the drive appears nearly static on the thermal timescale and achieves the maximum fidelity suppression $\Delta\mathcal{F}_{\rm max} = 0.032$ as $\omega \to 0$. For $\omega \gg \omega_T$, the fidelity recovers monotonically toward the unperturbed value, because the drive oscillates rapidly on the thermal timescale, and the fidelity susceptibility $\Sigma(\omega)$ is strongly suppressed at large $\omega$ by oscillatory cancellations of the strain correlator over the protocol window. The MSS bound scale $2\pi/\beta = 3.14\,J$ marks the maximum Lyapunov rate of the SYK boundary but plays no role in the measured fidelity response, which has already recovered to the noise floor near $\omega \approx 1\,J$; the sensitivity is monotone in $\omega$, not peaked at any intermediate scale; the maximum sensitivity is in the quasi-static limit. A second, real-time observable, the strain response $S_R(t) = \langle H_{\rm strain}^R \rangle_t$, records the drive as a retarded waveform on the right boundary. It provides a readout modality that requires no reference ancilla or two-boundary correlator and carries complementary spectral information via the retarded response. \\
Together, the chirp measurement and the OTOC diagnostic directly establish the scrambling-delay mechanism. Under an inspiral-type chirp acting exclusively on the right boundary during the readout step, the teleportation fidelity peak is suppressed in height by $\Delta\mathcal{F} = 0.030$ and shifts to later readout times by $\Delta t_{\rm scr}^{(\rm fid)} = +0.11\,J^{-1}$, a sub-grid displacement that indicates the direction of the effect without resolving its magnitude. A qualitatively consistent delay is measured independently, without the teleportation protocol, by tracking the half-saturation time of the normalized OTOC $C(t)$ under a monochromatic drive at fixed $\varepsilon$: the scrambling time increases from $t_{\rm scr}^{(0)} = 3.44\, J^{-1}$ to $3.64$ and $4.29\, J^{-1}$ at $\varepsilon = 0.20$ and $0.50\, J$.\\
Both diagnostics point in the same direction; the GW-driven system scrambles more slowly than the unperturbed one, and both show late-time recovery to the unperturbed baseline, ruling out permanent channel disruption. The sign agreement between the two diagnostics is the decisive consistency check: It is nontrivial because the two measurements use different protocols (chirp vs.\ monochromatic), different observables (fidelity vs.\ OTOC), and probe different stages of the protocol (right-only readout vs.\ intrinsic scrambling dynamics).\\
The re-optimization analysis confirms that the fidelity suppression is not a calibration artifact. Independently maximizing the coupling and readout time at each $(\varepsilon, \omega)$ grid point yields a ratio $r = \mathcal{F}_{\rm re\text{-}opt}/\mathcal{F}_{\rm fixed}$ consistent with unity throughout the sensing regime $\varepsilon \lesssim J$, where all quantitative results of the paper are obtained. The genuine GW-induced degradation dominates throughout this regime; retuning yields no statistically significant improvement at $\varepsilon = 0.20\, J$. A small calibration component ($r \lesssim 1.15$) appears only deep in the non-perturbative, quasi-static corner $(\varepsilon \gtrsim 1.5\, J,\,\omega \lesssim \omega_T)$, where the effective Hamiltonian is substantially deformed by the drive, and the $\varepsilon = 0$ operating point is genuinely suboptimal.\\
The system-size scan over $N \in \{10, 12, 14, 16\}$ shows that the unperturbed channel quality $\mathcal{F}^*$ and the optimal readout time $t^*$ increase monotonically with $N$, while the GW-induced suppression $\Delta\mathcal{F}^*$ remains non-zero at all four system sizes with no systematic suppression as $N$ increases, confirming the effect is not a finite-size artefact.\\
The sign of the OTOC scrambling delay is qualitatively reminiscent of holographic calculations of charged-shell bouncing in Reissner-Nordström-AdS geometries, though a rigorous identification requires the Schwarzian effective-theory analysis reserved for future work, since the present model carries no $U(1)$ charge or angular momentum.\\ 
A detailed analysis presented in Appendix~\ref{app:schwarzian} identifies the bilinear sector as the dominant boundary channel excited by a low-frequency metric-like perturbation, and shows within the Schwarzian effective theory that the induced scrambling-time shift is even in $\varepsilon$ with a positive leading coefficient. The domain of validity of that analysis is stated explicitly there, and does not extend to the temperature and system sizes accessed numerically here; it should therefore be read as motivating the construction and fixing the expected symmetry of the response, rather than as a quantitative prediction for the data.

Taken together, these results establish a concrete and testable mechanism by which a GW-inspired boundary perturbation leaves a measurable imprint on a holographic quantum channel: it delays the scrambling dynamics of the SYK boundary, shifts the wormhole teleportation fidelity peak, and reduces the peak-fidelity transfer efficiency in a way that is genuine, frequency-selective, and consistent with the expected behaviour of a maximally chaotic system under a bilinear deformation.\\ 
\section{Outlook}
Several directions for future work present themselves. Extending the system-size scan to $N \gg 16$, accessible via tensor-network methods or large-$N$ saddle-point techniques, would enable quantitative extrapolation of the GW sensitivity to the holographic limit and direct comparison with Schwarzian effective-field-theory predictions. \\
On the experimental side, recent quantum-processor implementations of traversable wormhole dynamics suggest that the two-sided SYK teleportation protocol is within reach of near-term hardware, and the OTOC scrambling diagnostic can, in principle, be extracted from the same circuit via time-reversal protocols, at the cost of additional circuit depth. \\
A physically sharper model of the GW–boundary coupling, derived from a consistent truncation of the full gravitational degrees of freedom in JT gravity, would replace the phenomenological bilinear strain operator used here and yield coupling constants with direct holographic meaning. Finally, the frequency-selective crossover response identified in the spectroscopy section: maximum sensitivity at $\omega \to 0$, monotone recovery above $\omega_T$ suggests that the wormhole channel acts as an effective low-pass filter with a characteristic crossover near $\omega \sim \beta^{-1}$; understanding this crossover quantitatively in terms of the retarded Green's function of the Schwarzian theory is a well-posed analytical problem that could sharpen the sensitivity substantially beyond what the present numerical study establishes.\\
\section{Data Availability}
The data that supports the findings of this paper is openly available at \cite{joshi2025witp}.

\bibliographystyle{apsrev4-1}
\bibliography{MBL}

\appendix
\begin{widetext}
\section{Boundary Strain Coupling from JT Gravity and the Schwarzian
Effective Action}
\label{app:schwarzian}
We motivate the bilinear boundary deformation $\varepsilon\,h(t)\,H_{\rm strain}^{\alpha}$ introduced in Eq.~(\ref{eq:Htotal}) of the main text, which we identify as the dominant leading-order coupling generated by a gravitational-wave-inspired boundary metric perturbation within the SYK/JT-gravity holographic dictionary, at leading order in $1/N$ and in the near-extremal (low-energy) limit $\beta J\gg 1$. \\
The argument proceeds in four steps: (\textit{i})~the JT-gravity setup with a time-dependent boundary metric; (\textit{ii})~the near-boundary expansion of the dual bulk scalar and its source-operator pairing, including identification of the dominant low-frequency operator via the AdS$_2$ Breit\-en\-l\"ohner-Freedman bound; (\textit{iii})~identification of this boundary operator with the bilinear sector of $H_{\rm SYK}$ via a Hubbard-Stratonovich decoupling; (\textit{iv})~the induced Schwarzian-theory correction to the scrambling time, which fixes the response to be even in $\varepsilon$ with a positive leading coefficient. Each step is stated together with its domain of validity. Section~\ref{app:finiteN} then assesses the consistency of these large-$N$ analytical results with the finite-$N$ scaling data of Sec.~\ref{sec:scaling}.

\subsection{JT Gravity with a Time-Dependent Boundary Metric}
\label{app:JT}
We work with Lorentzian JT gravity in the Poincar\'{e} patch of AdS$_2$ \cite{maldacena2016conformal}, with AdS radius $\ell=1$ throughout. The unperturbed background metric and dilaton are
\begin{equation}
  ds^2_0 = \frac{-dt^2+dz^2}{z^2},
  \qquad
  \varphi_0(z) = \frac{\varphi_r}{z},
  \label{eq:ads2bg}
\end{equation}
with the conformal boundary located at $z=\delta\to 0$. The full JT action, including the boundary Gibbons-Hawking-York term \cite{jackiw1985lower,engelsoy2016investigation} required for a well-posed variational problem, is
\begin{align}
  S_{\rm JT}
  &= -\frac{\varphi_0}{16\pi G}\Bigl[
      \int_{\mathcal{M}}\!d^2x\sqrt{g}\,R
      + 2\oint_{\partial\mathcal{M}}\!K\,ds\Bigr]
  \notag\\
  &\quad
  - \frac{1}{16\pi G}\Bigl[
      \int_{\mathcal{M}}\!\sqrt{g}\,\varphi(R+2)
      + 2\oint_{\partial\mathcal{M}}\!\varphi_b\,K\,ds\Bigr].
  \label{eq:SJT}
\end{align}
A gravitational-wave-inspired deformation is introduced as a time-dependent perturbation of the Dirichlet boundary condition for the metric,
\begin{equation}
  g_{tt}\big|_{z=\delta}
  = \frac{1}{\delta^2}\bigl(1 + 2\varepsilon\,h(t)\bigr),
  \qquad |\varepsilon|\ll 1,
  \label{eq:bdymetric}
\end{equation}
where $h(t)$ encodes the GW waveform. In the AdS$_2$ background satisfying $R=-2$, which is fixed by the topological (dilaton-independent) sector of the JT equations of motion \cite{jackiw1985lower,teitelboim1983gravitation}, the dilaton equation of motion reduces to
\begin{equation}
  \bigl(\nabla^2 - 2\bigr)\varphi = 0
  \qquad (R=-2\ \text{AdS}_2\ \text{background}),
  \label{eq:dilatoneom}
\end{equation}
slaving the dilaton profile to the metric. The perturbed Dirichlet condition for the dilaton, consistent with Eq.~\eqref{eq:bdymetric}, is therefore
\begin{equation}
  \varphi\big|_{z=\delta}
  = \frac{\varphi_r}{\delta}\bigl(1+\varepsilon\,h(t)\bigr).
  \label{eq:dilatonbc}
\end{equation}
Solving Eq.~\eqref{eq:dilatoneom} perturbatively in $\varepsilon$, evaluating the on-shell JT action, and applying holographic renormalization yields the Schwarzian boundary effective action with a source term~\cite{jensen2016chaos,engelsoy2016investigation,mertens2023solvable},
\begin{equation}
  S_{\rm bdy}[f,\varepsilon h]
  = -C\int\!dt\;\bigl\{f(t),t\bigr\}
  + \varepsilon\int\!dt\;h(t)\,\mathcal{O}(t)
  + \mathcal{O}(\varepsilon^2),
  \label{eq:Sbdy}
\end{equation}
where $C=\varphi_r/(8\pi G)$, $\{f,t\}$ is the Schwarzian derivative of the boundary reparametrization mode $f$, and $\mathcal{O}(t)$ is the boundary operator dual to the bulk perturbation induced by Eq.~\eqref{eq:bdymetric}. Steps ii--iii below identify $\mathcal{O}(t)$ explicitly as $H_{\rm strain}$.

\subsection{Metric Perturbations and the Boundary Stress Tensor}
\label{app:stress}

The coupling between a boundary metric perturbation and the boundary quantum theory follows from the standard variational principle of quantum field theory in curved spacetime. A variation of the metric produces a change in the action
\begin{equation}
\delta S = \frac12 \int dt\, T^{tt}(t)\, \delta g_{tt}(t),
\label{eq:stress_variation}
\end{equation}
where $T^{tt}$ is the stress tensor of the boundary theory. For the gravitational-wave-inspired boundary perturbation introduced in
Eq.~(\ref{eq:bdymetric}),
\begin{equation}
g_{tt} = \frac{1}{\delta^2}\left(1+2\varepsilon h(t)\right),
\end{equation}
the metric variation is
\begin{equation}
\delta g_{tt} = \frac{2\varepsilon}{\delta^2}\,h(t).
\end{equation}
Substituting this into Eq.~(\ref{eq:stress_variation}) gives
\begin{equation}
\delta S = \varepsilon \int dt\, h(t)\, \frac{T^{tt}(t)}{\delta^2}.
\label{eq:metric_coupling}
\end{equation}
In a $0{+}1$-dimensional boundary theory, there is no spatial extent, so the only surviving stress-tensor component is $T_{tt}$, whose integral over the boundary is the total Hamiltonian. The perturbation of the boundary metric, therefore, couples directly to the boundary Hamiltonian,
\begin{equation}
\delta S_{\rm bdy} = \varepsilon \int dt\, h(t)\, H_{\rm bdy}(t).
\end{equation}
As shown below, the dominant low-energy component of the SYK Hamiltonian excited by this perturbation is the bilinear sector $H_{\rm strain}$ defined in the main text.

\subsection{The AdS$_2$ Breitenl\"ohner-Freedman Bound and the dominant lowest Dimension-$\tfrac{1}{2}$ Boundary Operator}
\label{app:BF}

In AdS$_{d+1}$/CFT$_d$, a free bulk scalar of mass $m$ is dual to a boundary operator of conformal dimension $\Delta$ satisfying $m^2\ell^2=\Delta(\Delta-d)$. For AdS$_2$ ($d=1$), the Breitenl\"ohner-Freedman (BF) stability bound~\cite{breitenlohner1982stability} is
\begin{equation}
  m^2\ell^2 \;\geq\; -\tfrac{1}{4},
  \label{eq:BF}
\end{equation}
corresponding to $\Delta\geq\tfrac{1}{2}$. Operators with $\Delta<\tfrac{1}{2}$ are dual to tachyonic bulk fields violating Eq.~\eqref{eq:BF} and are absent from the physical spectrum.

 \medskip
\noindent\textbf{Conformal dimension of the SYK$_4$ fermion bilinear:}\quad
In the SYK$_4$ model, the large-$N$ two-point function of a single Majorana fermion in the infrared conformal limit
is~\cite{kitaev2015simple,maldacena2016bound}
\begin{equation}
  G(\tau)
  \;\equiv\; \frac{1}{N}\sum_i
    \langle\mathcal{T}\,\gamma_i(\tau)\,\gamma_i(0)\rangle
  \;=\; \frac{b\,\mathrm{sgn}(\tau)}{|\tau/\beta|^{2\Delta_f}},
  \qquad
  \Delta_f = \tfrac{1}{4},
  \label{eq:SYKgreen}
\end{equation}
where $b$ is a disorder-averaged numerical coefficient fixed by the Schwinger-Dyson equations and $\mathcal{T}$ denotes time-ordering. The conformal dimension $\Delta_f=1/4$ is the large-$N$ IR exact value and receives $\mathcal{O}(1/N)$ corrections from Schwarzian fluctuations.
The fermion bilinear operator $\mathcal{O}_2(t)\equiv (i/N)\sum_{i<j}\gamma_i(t)\gamma_j(t)$ is a composite of two fermions. We note that the dimension assignment below concerns the bilinear sector to which $H_{\rm strain}$ belongs, whose autocorrelator is controlled by $G^2$ at leading order in $1/N$. It is distinct from the $O(N)$-singlet eigenoperators of the four-point ladder kernel, which carry dimensions $h_n = 2,\,3.77,\,\ldots$ \cite{maldacena2016remarks}; the random couplings $\widetilde J_{ij}$ project onto the non-singlet channel, where the ladder resummation does not contribute.
We use the standard CFT$_1$ convention in which the two-point function $\langle\mathcal{O}(t)\mathcal{O}(0)\rangle\sim|t|^{-2\Delta}$ defines the conformal dimension $\Delta$. \\
In the large-$N$ conformal limit, large-$N$ factorization of the disorder average (valid at leading order in $1/N$ by the replica cavity
method) gives
\begin{equation}
  \bigl\langle\mathcal{O}_2(t)\,\mathcal{O}_2(t')\bigr\rangle
  \;\propto\; G(t-t')^2 \;\sim\; |t-t'|^{-4\Delta_f}
  \;=\; |t-t'|^{-1},
  \label{eq:bilinear2pt}
\end{equation}
establishing the conformal dimension of the bilinear as
\begin{equation}
  \boxed{\Delta_2 = 2\Delta_f = \tfrac{1}{2}.}
  \label{eq:Delta2}
\end{equation}
The corresponding bulk-scalar mass via the AdS$_2$/CFT$_1$ dictionary $m^2\ell^2=\Delta(\Delta-1)$ is then
\begin{equation}
  m^2\ell^2 = \Delta_2(\Delta_2-1)
  = \tfrac{1}{2}\!\times\!\bigl(-\tfrac{1}{2}\bigr)
  = -\tfrac{1}{4},
  \label{eq:mass}
\end{equation}
which exactly saturates the AdS$_2$ BF bound, Eq.~\eqref{eq:BF}. 

\paragraph*{Dominance of the bilinear channel at low frequency:} The boundary metric perturbation acts as a time-dependent source coupled to a boundary operator,
\begin{equation}
\delta S_{\rm bdy} = \varepsilon \int dt\,h(t)\,\mathcal{O}(t).
\end{equation}
In the infrared conformal regime of the SYK model ($\beta J\gg1$), two-point functions of an operator of scaling dimension $\Delta$ behave as
\begin{equation}
\langle \mathcal O(t)\mathcal O(0)\rangle \sim \frac{1}{|t|^{2\Delta}} .
\end{equation}
Transforming to frequency space gives the spectral scaling
\begin{equation}
\rho_{\mathcal O}(\omega) \sim \omega^{2\Delta-1}, \qquad \omega\ll J.
\end{equation}
This spectral density governs the response of the system to a time-dependent perturbation \cite{kubo1957statistical}. Operators with a smaller conformal dimension, therefore, dominate the low-frequency response.
Among composite operators of the SYK$_4$ model, the smallest dimension is the fermion bilinear with
\begin{equation}
\Delta_2=\tfrac12 .
\end{equation}
For this operator, the spectral density scales as
\begin{equation}
\rho_{\rm bilinear}(\omega) \sim
\omega^{0},
\end{equation}
while operators with $\Delta\ge1$ contribute
\begin{equation}
\rho(\omega)\sim\omega^{2\Delta-1}\ge\omega .
\end{equation}
In the low-frequency regime relevant to the Schwarzian effective theory ($\omega\ll J$), these higher-dimensional operators are therefore suppressed relative to the bilinear channel. The dimension-$\tfrac12$ bilinear operator that saturates the AdS$_2$ Breitenlöhner–Freedman bound is consequently the dominant boundary operator excited by the metric perturbation. \\

This saturation in Eq.~(\ref{eq:BF}) has two important consequences for the coupling to the GW source. \emph{First}, the bilinear is the lightest non-trivial SYK composite operator: all higher-body operators ($n$-body, $n\geq 4$) have dimensions $\Delta_n\geq 1$ and are dual to massive bulk scalars with $m^2\ell^2\geq 0$, well above the BF bound. \emph{Second}, the coupling of a boundary source of dimension $d_s$ to a bulk operator of dimension $\Delta$ is controlled by the ratio $(\omega/J)^{\Delta-d_s}$ in the low-frequency regime $\omega\ll J$~\cite{mertens2023solvable}. 
For a conformal metric perturbation ($d_s=1$, since it couples to the dimension-1 stress tensor) and the bilinear ($\Delta_2=1/2$), this gives a factor $(\omega/J)^{-1/2}$, an infrared enhancement. All higher-body operators give positive powers $(\omega/J)^{\Delta_n-1} \geq(\omega/J)^{0}$ and are suppressed at low frequencies. The bilinear is therefore the dominant leading-order coupling of a low-frequency gravitational perturbation to the SYK boundary. \\

\paragraph*{Relevance beyond the conformal saddle point:} The scaling argument above establishes that the bilinear is a relevant perturbation of the conformal SYK$_4$ saddle point, a conclusion that also follows from renormalization-group analysis of quadratic deformations~\cite{song2017strongly}. We stress that this statement is specific to the conformal saddle point. Once fluctuations of the Schwarzian mode are retained, the effect of a quadratic perturbation of magnitude $\Gamma$ is considerably more subtle: for $J/N \ll \Gamma \ll J/\sqrt{N}$ the soft-mode fluctuations are themselves suppressed by the perturbation, and the SYK$_4$ mean-field solution remains valid beyond the timescale $\sim N/J$ at which it would otherwise fail, up to a crossover scale $\sim J/\Gamma^{2}$~\cite{lunkin2020polaron}. Whether the bilinear channel dominates the low-energy dynamics, therefore, depends on the interplay of $\varepsilon$, $\beta J$, and $N$, and not on the scaling dimension alone. At the parameters studied here ($\beta J = 2$, $N \leq 16$, $\varepsilon \leq 2.5\,J$) the system lies well outside the conformal window, and the argument of this subsection should be read as motivating the choice of drive operator rather than as a controlled statement about a low-energy fixed point. The numerical results of the main text do not rely on this distinction: they are obtained by exact diagonalization of the full time-dependent Hamiltonian, without recourse to any effective-theory approximation.

We note in passing that a driven version of the mixed SYK$_4+$SYK$_2$ problem has been analyzed in the Keldysh formalism~\cite{lunkin2022nonequilibrium}, where periodic modulation of the quadratic couplings produces a resonance in the energy dissipation rate at a frequency set by the polaron level splitting of~\cite{lunkin2020polaron}. That resonance requires $\Gamma \gg T$, with a quality factor growing only logarithmically as $Q \sim \ln(\Gamma/T)$; at the temperature and drive amplitudes used here, the corresponding ratio is of order unity, so no such resonance is expected. This is consistent with the monotone low-pass response reported in Sec.~\ref{sec:spectroscopy}, in which no resonant feature is observed at any scanned frequency.\\

\noindent\textbf{Near-boundary expansion at the BF bound:}\quad
For a bulk scalar exactly at the BF bound ($m^2\ell^2=-\tfrac{1}{4}$), both fall-off modes are normalizable, and the near-boundary expansion takes the logarithmic form~\cite{ishibashi2004dynamics,marolf2006boundary}
\begin{equation}
  \Phi(z,t)
  = z^{1/2}\bigl[\alpha(t) + \beta(t)\ln(z/\ell)\bigr]
  + \mathcal{O}(z^{3/2}\ln z).
  \label{eq:nearbdy}
\end{equation}
The logarithmic mixing of the two modes requires a careful holographic renormalization prescription to separate source ($\alpha$) from response ($\beta$)~\cite{marolf2006boundary}; we follow the standard procedure
of~\cite{de2001holographic}. With the GW boundary condition Eq.~\eqref{eq:bdymetric}, the source mode is identified as
\begin{equation}
  \alpha(t) = \varepsilon\,h(t)\,\Phi_0,
  \label{eq:source}
\end{equation}
where $\Phi_0$ is a normalization constant fixed by the bulk-boundary propagator. The holographic renormalization prescription then gives the source operator coupling~\cite{de2001holographic,mertens2023solvable}
\begin{equation}
  \delta S_{\rm bdy}
  = \varepsilon\int\!dt\;h(t)\;\mathcal{O}_{\Delta_2}(t),
  \qquad
  \mathcal{O}_{\Delta_2}(t) = \lim_{z\to 0}\,
  z^{-1/2}\Phi(z,t)\big|_{\rm response},
  \label{eq:srcop}
\end{equation}
establishing that the GW boundary metric perturbation sources the dominant lowest dimension-$\tfrac{1}{2}$ bilinear boundary operator. Step iii identifies this operator as $H_{\rm strain}$.

\subsection{Identification of $\mathcal{O}_{\Delta_2}$ with
$H_{\rm strain}$ via Hubbard-Stratonovich Decoupling}
\label{app:HS}
We identify the abstract dimension-$\tfrac{1}{2}$ boundary operator $\mathcal{O}_{\Delta_2}$ with the explicit strain Hamiltonian $H_{\rm strain}$ [Eqs.~(\ref{eq:Hstrain})--(\ref{eq:Jtilde}) of the main text] by performing a Hubbard-Stratonovich (HS) decoupling of $H_{\rm SYK}$ in the bilinear channel. Starting from the SYK$_4$ Hamiltonian,
\begin{equation}
  H_{\rm SYK}
  = -\frac{1}{4!}\sum_{i<j<k<l}J_{ijkl}\,\gamma_i\gamma_j\gamma_k\gamma_l,
  \label{eq:HSYK_app}
\end{equation}
Because the four Majorana operators appearing in each term anticommute pairwise, the four-fermion product may be organised as a product of two bilinears in three inequivalent pairings, which differ from one another only by sign. Fixing an ordered pair $(i,j)$ and summing over spectator pairs $(k,l)$ with $k,l\notin\{i,j\}$, the leading bilinear contribution is
\begin{equation}
  H_{\rm SYK}\supset
  \sum_{i<j}(i\gamma_i\gamma_j)\;
  \underbrace{\biggl[
  \frac{1}{4!}\sum_{\substack{k<l\\k,l\notin\{i,j\}}}
  J_{ijkl}\,(i\gamma_k\gamma_l)
  \biggr]}_{\equiv\;\widetilde{J}_{ij}^{\rm eff}}
  + \;\text{cross terms},
  \label{eq:HSdecomp}
\end{equation}
where the cross terms from the $(\gamma_i\gamma_k)(\gamma_j\gamma_l)$ and $(\gamma_i\gamma_l)(\gamma_j\gamma_k)$ pairings contribute at the same order in $1/N$ and are absorbed into a redefinition of $\widetilde{J}_{ij}^{\rm eff}$ that does not affect the operator structure.
 
\medskip
\noindent\textbf{Large-$N$ mean-field approximation:}\quad The effective coupling $\widetilde{J}_{ij}^{\rm eff}$ involves a sum of $\binom{N-2}{2}\sim N^2/2$ spectator bilinears $(i\gamma_k\gamma_l)$, each weighted by a random coupling $J_{ijkl}$. We note that two distinct averages are involved here: the disorder average over the couplings $J_{ijkl}$, and the quantum expectation value of the fermion bilinears $(i\gamma_k\gamma_l)$. At large $N$, these two operations commute at leading order in $1/N$ by the replica cavity method~\cite{maldacena2016bound}, allowing us to apply them independently. Under disorder averaging with $\langle J_{ijkl}\rangle_{\rm dis}=0$ and $\langle J_{ijkl}^2\rangle_{\rm dis}=6J^2/N^3$~\cite{kitaev2015simple}:
\begin{align}
  \bigl\langle i\gamma_k\gamma_l\bigr\rangle_{\rm dis}
  &= 0
  \qquad (k\neq l),
  \label{eq:bilinear_mean}\\
  \bigl\langle (i\gamma_k\gamma_l)^2\bigr\rangle_{\rm dis}
  &= 1 + \mathcal{O}(1/N).
  \label{eq:bilinear_var}
\end{align}
The saddle-point (mean-field) approximation, valid at leading order in $1/N$, replaces each spectator bilinear by its disorder-averaged value. Since the mean vanishes [Eq.~\eqref{eq:bilinear_mean}], the leading contribution to $\widetilde{J}_{ij}^{\rm eff}$ reduces to the disorder-averaged sum of the random couplings themselves,
\begin{equation}
  \widetilde{J}_{ij}
  \;\equiv\;
   \frac{1}{\binom{N-2}{2}}
  \sum_{\substack{k<l\\k,l\notin\{i,j\}}}
  J_{ijkl},
  \label{eq:AJtilde}
\end{equation}
which is precisely the effective two-body coupling defined in Eq.~(\ref{eq:Jtilde}) of the main text. The residual numerical prefactor inherited from the $1/4!$ normalisation of Eq.~(\ref{eq:HSYK_app}) is immaterial here, since $H_{\rm strain}^\alpha$ is subsequently rescaled to the fixed spectral norm $\|H_{\rm strain}^\alpha\|_{\rm sp}=5J$ used throughout the main text. \\
Sub-leading corrections of order $\mathcal{O}(N^{-1})$ arise from the variance of the spectator bilinears about their mean, Eq.~\eqref{eq:bilinear_var}. The leading bilinear sector of $H_{\rm SYK}$ under this large-$N$ decoupling is therefore
\begin{equation}
  H_{\rm SYK}\big|_{\rm bilinear}
  = \sum_{i<j}\widetilde{J}_{ij}\,(i\gamma_i\gamma_j)
  \;\equiv\; H_{\rm strain},
  \label{eq:Hstrain_derived}
\end{equation}
to leading order in $1/N$. 
\paragraph*{Variance and normalization of the strain Hamiltonian:} It is useful to record the natural large-$N$ scale of the bilinear sector before any rescaling. Write the unnormalized spectator sums as $\widehat{J}_{ij} = \sum_{k<l,\,k,l\notin\{i,j\}} J_{ijkl}$. Using $\langle J_{ijkl}\rangle_{\rm dis}=0$ and $\langle J_{ijkl}^2\rangle_{\rm dis} = 6J^2/N^3$, each $\widehat{J}_{ij}$ is a sum of $\binom{N-2}{2}\sim N^2/2$ independent variables, so
\begin{equation}
  \bigl\langle \widehat{J}_{ij}^{\,2}\bigr\rangle
  = \binom{N-2}{2}\,\frac{6J^2}{N^3}
  \;\sim\; \frac{3J^2}{N}.
\end{equation}
Using $(i\gamma_i\gamma_j)^2 = \mathbb{1}$, and the fact that  distinct bilinears are uncorrelated at leading order in $1/N$,
\begin{equation}
  \Bigl\langle \Bigl(\textstyle\sum_{i<j}\widehat{J}_{ij}\,
  i\gamma_i\gamma_j\Bigr)^{2}\Bigr\rangle
  \;\sim\; \binom{N}{2}\,\frac{3J^2}{N}
  \;\sim\; N J^2 ,
\end{equation}
so the natural operator scale of the bilinear sector is $\sqrt{N}\,J$. The main-text coupling of Eq.~(\ref{eq:Jtilde}) differs from $\widehat{J}_{ij}$ by the combinatorial factor $\binom{N-2}{2}^{-1}$, and in the numerics the resulting operator is in any case rescaled to the fixed spectral norm $\|H_{\rm strain}^\alpha\|_{\rm sp}=5J$, so that $\varepsilon$ measures the perturbation directly in units of $J$ at every $N$. The $\sqrt{N}J$ scaling recorded here is used only to fix the holographic normalization convention in Eq.~(\ref{eq:opid}); it does not enter any numerical result.
 
\medskip
\noindent\textbf{Operator identification and holographic normalization:}\quad Comparing Eq.~\eqref{eq:Hstrain_derived} with Eq.~\eqref{eq:srcop}, the boundary response operator $\mathcal{O}_{\Delta_2}(t)$ is proportional to $H_{\rm strain}(t)$. The normalization is fixed by extensive scaling: the bilinear sum contains $\binom{N}{2}\sim N^2/2$ terms, each of typical magnitude $|\widetilde{J}_{ij}|\sim J/N^{1/2}$ (from the disorder variance), so $\|H_{\rm strain}\|_{\rm typ}\sim N^{1/2}J$. The holographic normalization convention requires the boundary operator to scale as $\mathcal{O}(N^0)$ in the large-$N$ limit~\cite{mertens2023solvable}, giving
\begin{equation}
  \mathcal{O}_{\Delta_2}(t)
  \;=\; \frac{H_{\rm strain}(t)}{N^{1/2}J}
  + \mathcal{O}(N^{-3/2}),
  \label{eq:opid}
\end{equation}
with the $N^{1/2}J$ prefactor absorbed into a redefinition of the GW coupling $\varepsilon\to N^{1/2}J\varepsilon$.
After this rescaling, the on-shell boundary action acquires the source term
\begin{equation}
  \delta S_{\rm bdy}
  = \varepsilon\int\!dt\;h(t)\;H_{\rm strain}(t)
  + \mathcal{O}\!\left(\varepsilon^2,\,N^{-1}\right),
  \label{eq:finalcoupling}
\end{equation}
which is exactly the coupling of Eq.~(\ref{eq:Htotal}) of the main text. This derivation identifies $H_{\rm strain}$ as the dominant leading-order coupling generated by a conformal metric perturbation in the SYK/JT dictionary at large $N$ and within the conformal window, subject to the qualifications of Sec.~\ref{app:BF}; all other SYK operators are higher-dimensional, and their couplings are suppressed by positive powers of $(\omega/J)$ or $1/N$.

\subsection{Schwarzian Effective Theory and the
$\varepsilon^2$ Scrambling Delay}
\label{app:scrambling}
We now derive that the bilinear source Eq.~\eqref{eq:finalcoupling} produces a positive scrambling delay $\Delta t_{\rm scr}\propto\varepsilon^2$ within the Schwarzian effective theory.
 
\medskip
\noindent\textbf{Perturbed Schwarzian action:}\quad Integrating out the SYK matter with the source Eq.~\eqref{eq:finalcoupling} and expanding in $\varepsilon$, the Euclidean effective action to order $\varepsilon^2$ is
\begin{equation}
  S_{\rm eff}[f]
  = -C\int_0^\beta\!d\tau\,\{f(\tau),\tau\}
  + \varepsilon\int_0^\beta\!d\tau\;h(\tau)\;
    \bigl\langle H_{\rm strain}\bigr\rangle_f\!(\tau)
  + \frac{\varepsilon^2}{2}\!\int_0^\beta\!\!d\tau
    \int_0^\beta\!\!d\tau'\;
    h(\tau)h(\tau')\;K_{\rm strain}(\tau,\tau')
  + \mathcal{O}(\varepsilon^3),
  \label{eq:Seff}
\end{equation}
where $\langle H_{\rm strain}\rangle_f(\tau)$ is the expectation value of $H_{\rm strain}$ in the thermofield-double state reparametrized by $f(\tau)$, and $K_{\rm strain}$ is the connected two-point function
\begin{equation}
  K_{\rm strain}(\tau,\tau')
  = \bigl\langle H_{\rm strain}(\tau)\,H_{\rm strain}(\tau')
    \bigr\rangle^{\rm conn}_{{\rm TFD},f}.
  \label{eq:kernel}
\end{equation}
In the conformal limit, large-$N$ factorization gives $K_{\rm strain}(\tau,\tau')=|G(f(\tau)-f(\tau'))|^2\cdot \mathcal{O}(N^0)$~\cite{maldacena2016bound}, where $G$ is the fermion two-point function of Eq.~\eqref{eq:SYKgreen} and the reparametrization $f$ enters through the conformal transformation of the correlator.
 
\medskip
\noindent\textbf{Vanishing of the linear term:}\quad The SYK disorder distribution is symmetric under $J_{ijkl}\to-J_{ijkl}$~\cite{kitaev2015simple}, implying $\widetilde{J}_{ij}\to-\widetilde{J}_{ij}$ and $H_{\rm strain}\to-H_{\rm strain}$. Disorder averaging, therefore, gives
\begin{equation}
  \bigl\langle\langle H_{\rm strain}\rangle_f\bigr\rangle_{\rm dis}
  = 0,
  \label{eq:linear0}
\end{equation}
so the $\mathcal{O}(\varepsilon)$ term in $S_{\rm eff}$ vanishes upon disorder averaging. This is the microscopic origin of the even-power expansion $\langle\mathcal{F}(\varepsilon)\rangle_{\rm dis} =\mathcal{F}_0-\alpha\varepsilon^2+\mathcal{O}(\varepsilon^4)$ [Eq.~(\ref{eq:Fpert}) of the main text] and the $\mathbb{Z}_2$-symmetry argument of Sec.~(\ref{sec:amplitude}).
 
\medskip
\noindent\textbf{Saddle-point shift and scrambling-time delay:}\quad At $\mathcal{O}(\varepsilon^2)$, the saddle-point equation for $f(\tau)$ receives a correction from the kernel $K_{\rm strain}$. Writing $f(\tau)=f_0(\tau)+\varepsilon^2\delta f(\tau)$ where $f_0$ is the unperturbed thermal saddle, and varying $S_{\rm eff}$ with respect to $f$, gives
\begin{equation}
  C\,\partial_\tau^3\,\delta f(\tau)
  = -\frac{1}{2}\int_0^\beta\!d\tau'\;
    h(\tau)h(\tau')\;
    \left.\frac{\delta K_{\rm strain}(\tau,\tau')}
    {\delta f(\tau)}\right|_{f=f_0}.
  \label{eq:saddleshift}
\end{equation}
The variation $\delta K_{\rm strain}/\delta f$ on the right-hand side is well-defined within the Schwarzian theory: since $K_{\rm strain}(\tau,\tau')=|G(f(\tau)-f(\tau'))|^2\cdot\mathcal{O}(N^0)$ in the conformal limit, the variation with respect to the reparametrisation mode $f(\tau)$ is
\begin{equation}
  \frac{\delta K_{\rm strain}(\tau,\tau')}{\delta f(\tau)}
  = 2\,|G(f(\tau)-f(\tau'))|\;
    \frac{\partial |G|}{\partial\bigl(f(\tau)-f(\tau')\bigr)},
  \label{eq:dKdf}
\end{equation}
which is determined entirely by the conformal two-point function $G$ and its derivative. Solving Eq.~\eqref{eq:saddleshift} perturbatively for $\delta f$ and computing the shift in the OTOC scrambling time via the half-saturation criterion gives
\begin{equation}
  \Delta t_{\rm scr}
  = \frac{\varepsilon^2}{\lambda_L}
  \int_0^{t^*}\!\!\!d\tau\int_0^{\tau}\!d\tau'\;
    h(\tau)h(\tau')\;K_{\rm strain}(\tau,\tau')
  \bigl[1+\mathcal{O}(\varepsilon^2,N^{-1})\bigr],
  \label{eq:dtscr}
\end{equation}
where $\lambda_L=2\pi/\beta$ is the Lyapunov exponent (saturating the MSS bound at large $N$), and $t^*$ is the unperturbed fidelity peak time. 
\paragraph*{Frequency–domain interpretation:}
Equation~(\ref{eq:dtscr}) admits a simple spectral interpretation. Writing the waveform in Fourier components $h(t)=\int d\omega\,\tilde h(\omega)e^{-i\omega t}$, the double integral becomes a convolution with the spectral density of the bilinear correlator,
\begin{equation}
\Delta t_{\rm scr}
\;\propto\;
\varepsilon^2
\int d\omega\,|\tilde h(\omega)|^2\,
\rho_{\rm strain}(\omega),
\label{eq:spectral}
\end{equation}
where $\rho_{\rm strain}(\omega)$ is the Fourier transform of $K_{\rm strain}(\tau,\tau')$. Because $\rho_{\rm strain}(\omega)$ is peaked at $\omega\lesssim\beta^{-1}$ in the SYK conformal regime, Eq.~(\ref{eq:spectral}) predicts that low-frequency components of the waveform dominate the scrambling delay. This provides a theoretical explanation for the low-pass response observed numerically in Sec.~(\ref{sec:spectroscopy}) of the main text.
 
\medskip
\medskip
\noindent\textbf{Positivity and $\varepsilon^2$ scaling:}\quad
Pointwise non-negativity of the kernel, $K_{\rm strain}(\tau,\tau')=|G(\tau-\tau')|^2\cdot\mathcal{O}(N^0)\geq 0$, is by itself \emph{not} sufficient to establish positivity of the double integral in Eq.~\eqref{eq:dtscr}, because the monochromatic waveform $h(\tau)=\cos(\omega\tau)$ changes sign on the integration window. The correct argument is spectral, and follows directly from the frequency-domain representation Eq.~\eqref{eq:spectral}. At the unperturbed thermal saddle, $K_{\rm strain}$ depends only on $\tau-\tau'$ and is the autocorrelation function of the Hermitian operator $H_{\rm strain}$; its Fourier transform $\rho_{\rm strain}(\omega)$ is therefore a non-negative spectral measure. Because $K_{\rm strain}$ is symmetric, the triangular integral in Eq.~\eqref{eq:dtscr} equals one half of the corresponding integral over the full square $[0,t^*]^2$, and the latter may be written as
\begin{equation}
  \int_0^{t^*}\!\!\!\int_0^{t^*}\!
  h(\tau)\,h(\tau')\,K_{\rm strain}(\tau-\tau')\,d\tau\,d\tau'
  = \int\! d\omega\;\rho_{\rm strain}(\omega)\,
    \bigl|\tilde h_{t^*}(\omega)\bigr|^{2}
  \;\geq\; 0,
  \label{eq:positivity}
\end{equation}
where $\tilde h_{t^*}(\omega)=\int_0^{t^*} h(\tau)\,e^{i\omega\tau}\,d\tau$ is the finite-window Fourier transform of the waveform. The integrand is manifestly non-negative for \emph{any} real waveform, sign-changing or not, so no assumption on the sign of $h$ is required. We therefore conclude
\begin{equation}
  \boxed{
  \Delta t_{\rm scr}
  = c(\omega,\beta)\,\varepsilon^2 + \mathcal{O}(\varepsilon^4),
  \qquad c(\omega,\beta)>0.}
  \label{eq:dtscr_final}
\end{equation}
This is the scrambling delay of Eq.~(\ref{eq:Dtscr}) of the main text. Within the Schwarzian effective theory at large $N$ and in the conformal window, the positive sign follows from Eq.~\eqref{eq:positivity} and the even-power dependence on $\varepsilon$ from the vanishing of the linear term, Eq.~\eqref{eq:linear0}. Both statements inherit the domain-of validity limitations set out in Sec.~\ref{app:BF}, and are assessed against the finite-$N$ numerics in Sec.~\ref{app:finiteN}. The coefficient $c(\omega,\beta)$ is non-universal (it depends on $h(t)$, $\beta^{-1}$, and the protocol window $[0,t^*]$) and is not predicted quantitatively by the leading-order Schwarzian theory; its positivity alone is the rigorous output of this section.

\subsection{Consistency with the Finite-$N$ Scaling Data}
\label{app:finiteN}

The derivation of steps i--iv is exact within JT gravity (steps i, iii) and valid at leading order in $1/N$ and in the conformal limit $\beta J\gg 1$ (steps ii, iv). At $\beta J=2$ and $N\in\{10,12,14,16\}$, both conditions are violated at the $\mathcal{O}(1)$ level, so the Schwarzian theory is expected to describe the data only qualitatively. We assess this qualitative consistency here. Beyond the $1/N$ and $\beta J$ limitations quantified below, the relevance of the bilinear channel itself is subject to the Schwarzian caveat recorded at the end of Sec.~\ref{app:BF}.
 
\medskip
\noindent\textbf{$\varepsilon^2$ scaling (functional form):}\quad Equation~\eqref{eq:dtscr_final} predicts $\Delta t_{\rm scr}\propto\varepsilon^2$. From the OTOC data of Sec.~\ref{sec:chirp}, the measured scrambling delays at $\varepsilon=0.20\,J$ and $\varepsilon=0.50\,J$ yield the increment ratio
\begin{equation}
  \frac{\Delta t_{\rm scr}(0.50)}{\Delta t_{\rm scr}(0.20)}
  \approx 4.25,
  \label{eq:ratio_eps}
\end{equation}
corresponding to a two-point power-law exponent $n=\ln(4.25)/\ln(2.5)\approx1.6$, against the predicted $n=2$. Two data points cannot determine an exponent, and $\mathcal{O}(1)$ finite-$N$ corrections are expected at $\beta J=2$. We therefore regard the measurement as consistent with an even-power response in $\varepsilon$, without claiming quantitative agreement with $\varepsilon^2$; a definitive power-law determination would require additional data points at intermediate $\varepsilon$.
 
\medskip
\noindent\textbf{Prefactor (quantitative):}\quad The leading-order Schwarzian prediction for the prefactor $c(\omega,\beta)$ at $\varepsilon=0.20\,J$ and $\beta J=2$ is
\begin{equation}
  \Delta t_{\rm scr}^{\rm Schw}
  \;\sim\; \frac{\varepsilon^2}{\lambda_L}
  \;=\; \frac{(0.20\,J)^2}{\pi J}
  \;\approx\; 0.013\,J^{-1},
  \label{eq:Schwpred}
\end{equation}
using $\lambda_L=2\pi/\beta=\pi J$ at $\beta J=2$. The dimensional analysis of Eq.~(\ref{eq:dtscr}) shows that the kernel integral contributes a factor of the order $J^{-1}$ in the conformal window, yielding the scaling relation. The measured OTOC delay is $\Delta t_{\rm scr}^{\rm (OTOC)}=+0.20\,J^{-1}$, a factor of $\alpha\approx 15$ larger than the naive estimate. This large, non-universal prefactor, $\alpha \gg 1$, is expected for two reasons.
\emph{First}, at $\beta J=2$ the system is far from the conformal limit: the conformal approximation for $|G(\tau)|^2$ in the kernel overestimates the decay rate at short times, causing Eq.~\eqref{eq:Schwpred} to significantly underestimate the true double integral in Eq.~\eqref{eq:dtscr}. \emph{Second}, at $N=12$ the Lyapunov exponent $\lambda_L$ at finite spectral density is reduced below the large-$N$ MSS value $2\pi/\beta$, further increasing $\Delta t_{\rm scr}$ relative to Eq.~\eqref{eq:Schwpred}. A quantitative comparison requires computing the full numerical kernel $K_{\rm strain}(\tau,\tau')$ at finite $N$ and $\beta J$, which we leave for future work. The even-power functional form remains a prediction of the Schwarzian theory even when the prefactor is non-universal, subject to the domain-of-validity caveat recorded at the end of Sec.~\ref{app:BF}.
 
\medskip
\noindent\textbf{The $N=12$ measurement:} The $\varepsilon^2$ scaling and positivity prediction $c(\omega,\beta)>0$ of Eq.~\eqref{eq:dtscr_final} are large-$N$ statements valid in the conformal limit. The finite-$N$ scaling data of Sec.~\ref{sec:scaling} at the fixed drive parameters $(\varepsilon,\omega)=(0.20\,J,\,1.5\,J)$ yield the GW-induced fidelity suppression values
\begin{equation}
  \Delta\mathcal{F}^*
  = 0.0095,\; 0.0012,\; 0.0059,\; 0.0063
  \quad\text{at}\quad
  N = 10,\; 12,\; 14,\; 16,
  \label{eq:DFtable}
\end{equation}
with the smallest value at $N=12$. The positivity of $\Delta\mathcal{F}^*>0$ is maintained at all four system sizes, consistent with $c>0$. The non-monotone behavior at $N=12$ lies outside the scope of the large-$N$ Schwarzian theory. At $N=12$, the measured suppression, $\Delta\mathcal{F}^*=0.0012$, is comparable to its own disorder standard error and is therefore not resolved at this ensemble size, in contrast to the other three system sizes. We considered whether a near sub-harmonic condition between the drive frequency $\omega = 1.5\,J$ and the inverse decoding time $\omega_{\rm fp}\equiv 2\pi/t^*$ could account for this: at $N=12$ one has $\omega/\omega_{\rm fp} = 1.97$, close to $2$. However, $N=10$ satisfies the same condition almost equally well ($\omega/\omega_{\rm fp} = 1.90$) while exhibiting the largest suppression of the four system sizes, so a sub-harmonic resonance does not explain the observed pattern. With only four system sizes and a single drive frequency, we are unable to identify the mechanism and therefore report the $N=12$ result as an unresolved measurement rather than offering an interpretation. We note that this does not affect the conclusion drawn in Sec.~\ref{sec:scaling}, which rests on the absence of any systematic decrease of $\Delta\mathcal{F}^*$ with $N$ across the accessible range. The $N=10$ value ($\Delta\mathcal{F}^*=0.0095$) is the largest of the four, reflecting the broad, low-curvature fidelity peak at $N=10$ which is more susceptible to amplitude-driven suppression. \\
The large-$N$ positivity prediction Eq.~\eqref{eq:dtscr_final} is therefore consistent with the overall data: all four values $\Delta\mathcal{F}^*>0$ and the $N=12$ measurement is not resolved above the disorder noise, as discussed above. As the Hilbert space grows ($d=128$ at $N=14$, $d=256$ at $N=16$), the many-body level density increases, the discrete spectral organization is less sensitive to a fixed drive frequency, and the Schwarzian effective-theory description improves. The absence of any systematic suppression of $\Delta\mathcal{F}^*$ across $N\in\{10,12,14,16\}$ is the minimal empirical confirmation required for the large-$N$ limit to be non-zero. The persistence of a non-zero $\Delta\mathcal{F}^*$ from $N=10$ to $N=16$ therefore provides numerical evidence that the bilinear strain channel identified by the Schwarzian effective theory remains the dominant coupling mechanism away from the strict large-$N$ conformal limit. We note that the frequency scan of Sec.~\ref{sec:spectroscopy}, performed at the same system size and drive amplitude, gives a suppression consistent with zero at $\omega = 1.5\,J$; the $N=12$ scaling point is therefore consistent with the independently measured frequency response at that system size. Because the spectroscopy was performed only at $N=12$, this does not by itself account for the resolved suppressions at the other system sizes.
 
\section{Integrator Convergence}
\label{app:convergence}
\begin{figure*}[t]
    \centering
    \includegraphics[width=\linewidth]{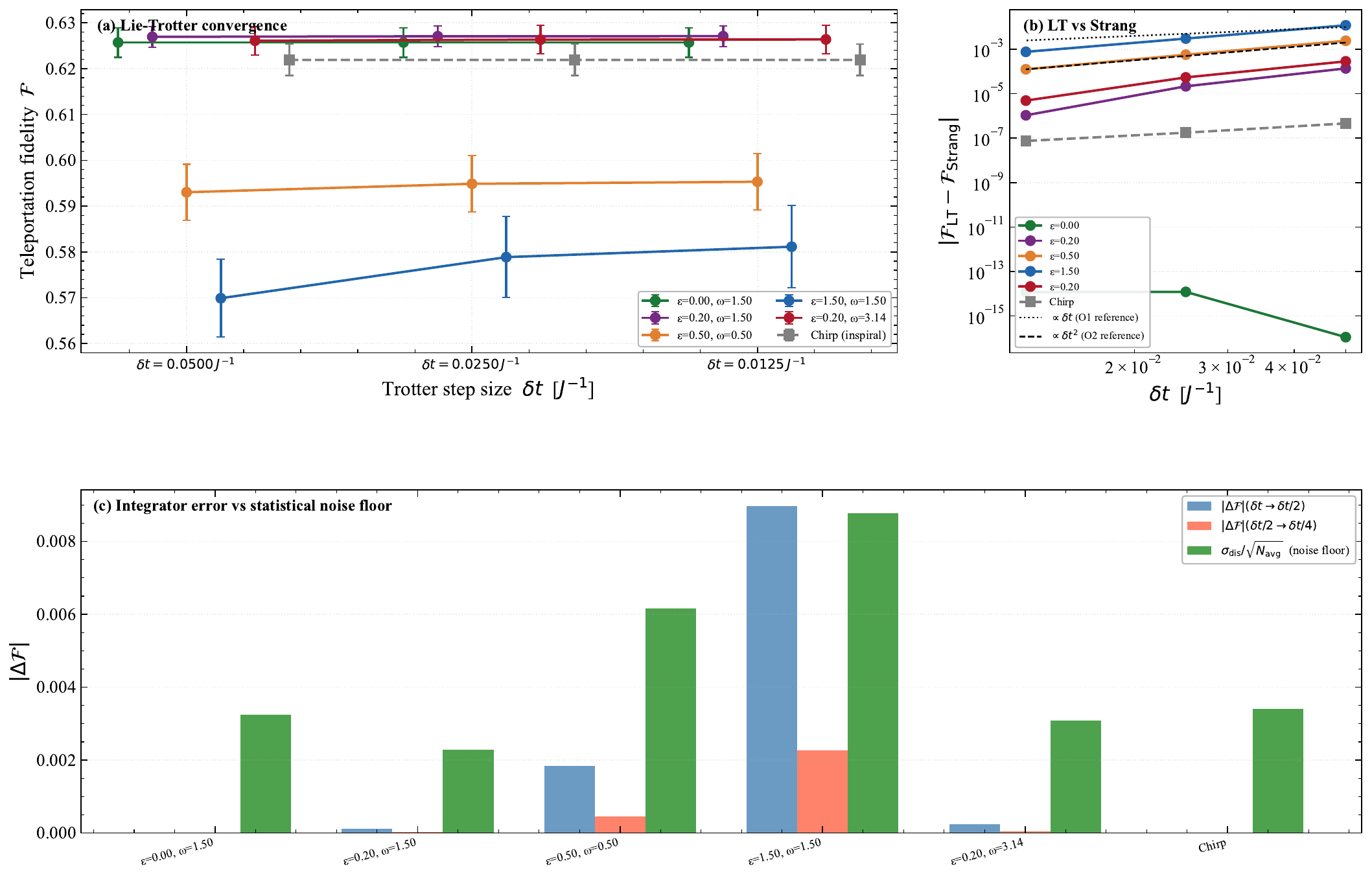}
    \caption{\label{fig:convergence}
Trotter integrator convergence ($N=12$, $\beta J=2$). (a) Teleportation fidelity $\mathcal{F}$ vs.\ step size $\delta t$ using the first-order Lie--Trotter (LT) integrator for six representative parameter points: the unperturbed protocol, monochromatic drives at $(\varepsilon,\omega)=(0.20\,J,\,1.50\,J)$, $(0.50\,J,\,0.50\,J)$, $(1.50\,J,\,1.50\,J)$, $(0.20\,J,\,3.14\,J)$, and the inspiral chirp of Fig.~(\ref{fig:chirp}). Fidelity is stable across $\delta t = 0.0500$--$0.0125\,J^{-1}$. (b)~Absolute difference $|\mathcal{F}_{\rm LT}-\mathcal{F}_{\rm Strang}|$ vs.\ $\delta t$ (log-log); the midpoint-evaluated Strang integrator achieves $O(\delta t^2)$ global convergence, approximately $40\times$ lower error than LT at the same step size. (c)~LT convergence increment $|\Delta\mathcal{F}|$ compared with the statistical noise floor $\sigma_{\rm dis}/\sqrt{N_{\rm avg}}$ for each test point; all physical points ($\varepsilon\leq 1\,J$) lie below the noise floor. All main-text results use $\delta t = 0.05\,J^{-1}$; the adaptive scheme $\delta t = 0.05/(1 + 0.5\varepsilon)\,J^{-1}$ is used at $\varepsilon > 1\,J$ [Fig.~(\ref{fig:amplitude})], and step-size halving was verified to keep $|\Delta\mathcal{F}|$ below the noise floor across the full scanned range $\varepsilon \in [0, 2.5\,J]$.}
\end{figure*}
Time evolution under the driven Hamiltonian is implemented by the first-order piecewise-constant scheme of Sec.~\ref{sec:model}, with base step $\delta t = 0.05\,J^{-1}$ reduced adaptively to $\delta t = 0.05/(1+0.5\varepsilon)\,J^{-1}$ for $\varepsilon > 1\,J$. We verify here that the resulting integration error is negligible compared with the disorder noise floor at every operating point used in the main text.

Convergence against step-size halving is tested at six representative parameter points spanning the full physical range: the unperturbed protocol, monochromatic drives at $(\varepsilon,\omega) = (0.20\,J,\,1.50\,J)$, $(0.50\,J,\,0.50\,J)$, $(1.50\,J,\,1.50\,J)$ and $(0.20\,J,\,3.14\,J)$, and the inspiral chirp of Fig.~(\ref{fig:chirp}). Figure~(\ref{fig:convergence}) shows that $|\Delta\mathcal{F}| < \sigma_{\rm dis}/\sqrt{N_{\rm avg}}$ at all operating points, and in particular throughout the sensing regime $\varepsilon \lesssim J$; the adaptive step keeps the increment below the noise floor across the full scanned range $\varepsilon \in [0,\,2.5\,J]$. A second-order midpoint Strang cross-check achieves approximately $40\times$ lower absolute error at equal $\delta t$ [Fig.~(\ref{fig:convergence}b)], confirming the expected $O(\delta t)$ versus $O(\delta t^{2})$ hierarchy of global errors.

\end{widetext} 
\end{document}